\documentclass[12pt]{article}

\usepackage{cite}
\usepackage{amsmath}
\usepackage{amssymb}
\usepackage{amsfonts}
\usepackage{wasysym}
\usepackage{cite}

\usepackage[dvips]{graphicx}
\DeclareGraphicsExtensions{.eps}

\newcommand{\rd}{{\rm d}}
\newcommand{\re}{{\rm e}}
\newcommand{\ri}{{\rm i}}

\newcommand{\be}{\begin{equation}}
\newcommand{\ee}{\end{equation}}
\newcommand{\bea}{\begin{eqnarray}}
\newcommand{\eea}{\end{eqnarray}}
\newcommand{\nn}{\nonumber}

\newcommand{\leapprox}{\apprle}

\newcommand{\ket}[1]{|#1\rangle}

\renewcommand{\vec}[1]{\mathbf{#1}}

\begin{document}

\title{Nonlinear Landau-Zener tunneling in quantum phase space}

\author{F Trimborn$^{1,2,3}$, D Witthaut$^{2,3,4}$, V Kegel$^3$, 
     and H J Korsch$^3$\\    
          \scriptsize
          ${}^1$ Institut f\"ur theoretische Physik, Leibniz Universit\"at Hannover, 
          D--30167 Hannover, Germany\\
             \scriptsize
          ${}^2$ QUANTOP, Niels Bohr Institute, University of Copenhagen, 
          DK--2100 Copenhagen, Denmark\\
             \scriptsize
           ${}^3$ Fachbereich Physik, TU Kaiserslautern,
          D--67663 Kaiserslautern, Germany\\
            \scriptsize
          ${}^4$ Max-Planck-Institute for Dynamics and Self-Organization, 
          D--37073 G\"ottingen, Germany
          }

\maketitle

\begin{abstract}
We present a detailed analysis of the Landau-Zener problem for an interacting Bose-Einstein condensate in a time-varying double-well trap, especially focussing on the relation between the full many-particle problem and the mean-field approximation. 
Due to the nonlinear self-interaction a dynamical instability occurs, which leads to a breakdown of adiabaticity condition and thus fundamentally alters the dynamics.
It is shown that essentially all features of the Landau-Zener problem including the depletion of the condensate mode can be already understood within a semiclassical phase space picture. In particular, this treatment resolves the formerly imputed incommutability of the adiabatic and semiclassical limits. The possibility to exploit Landau-Zener sweeps to generate squeezed states for spectroscopic tasks is analysed in detail. 
Moreover, we study the influence of phase noise and propose a Landau-Zener sweep as a sensitive, yet readily implementable probe for decoherence, since this has a significant effect on the transition rate for slow parameter variations.
\end{abstract}

\section{Introduction}
\label{sec-intro}

The Landau-Zener problem aims at the general description of nonadiabatic transitions at avoided level crossings. In the standard setting, the dynamics is restricted to two levels with a constant coupling $J$, whose energy difference varies linearly in time, $\epsilon(t)=\alpha t$. Of particular interest is the Landau-Zener tunneling probability between the two adiabatic states which is found to be
\begin{eqnarray} \label{eqn-plz-lin}
  P_{\rm LZ} = \re^{-\pi J^2/\alpha},
\end{eqnarray}
independent of the initially occupied level. Due to its generality, this result has been applied to numerous problems in various contexts like, e.g., spin-flip processes in nano-scale systems \cite{Wern99}, molecular collisions \cite{Chil74a}, quantum-dot arrays \cite{Spre90}, dissipative systems \cite{Wubs06,Kohl08} or quantum information processing tasks \cite{Coop04,Ithi05,Oliv05,Sait06}, to name but a few examples.

The Landau-Zener szenario was one of the first major problems adressed within time-dependant quantum theory. While the single-particle case was solved independently by Landau, Zener, Majorana and St\"uckelberg already in 1932 \cite{Land32b,Zene32,Majo32, Stue32}, the generalization of the results to interacting many-particle systems remains an open question up to today, and even the mean-field dynamics is not yet fully understood.  The non-linear self-interaction fundamentally alters the dynamics, leading to a breakdown of adiabaticity due to the bifurcation of nonlinear stationary states \cite{Wu00,Zoba00,Liu02,Wu03,05level3,06zener_bec,06bloch_manip}. 
The many-particle Landau-Zener problem is of fundamental interest not only from the theoretical but also from the experimental point of view and has in recent years attracted a lot of interest, especially in the context of the dynamics of Bose-Einstein condensates (BECs) in optical lattices \cite{Lasi03,Fall04,Sias07,Salg07,Zene09}.

In the following, we present a detailed analysis of nonlinear Landau-Zener tunneling between two modes focussing on the relation between the original full many-particle problem and the mean-field approximation. The breakdown of adiabaticity is a consequence of a bifurcation of the mean-field stationary states or the occurrence of near-degenerate avoided crossings in the many-particle spectrum, which are intimately related.
Furthermore we discuss the Landau-Zener problem within a semiclassical phase space picture, where
the quantum dynamics is approximated by a Liouvillian flow rather than a single trajectory. It is shown that  essentially all features of the dynamics including the depletion of the condensate mode can be already understood within this approach.
In particular, this treatment resolves the formerly imputed incommutability of the adiabatic and semiclassical limits. Number squeezing effects during the transition are analysed in detail. 
Moreover, we study the influence of phase noise which is an unavoidable feature in every experiment. Since it has a significant effect on the transition rate for slow parameter variations, a Landau-Zener sweep is a sensitive probe for decoherence. 

In particular, this paper is organized as follows: In section \ref{sec-mpmf} we first introduce the many-particle and mean-field description of the system and define the Landau-Zener transition probability $P_{\rm LZ}$ in both cases. The basic features of the nonlinear Landau-Zener problem are reviewed in section \ref{sec-lz1}. In the mean-field approximation, $P_{\rm LZ}$ does not vanish even in the limit $\alpha \rightarrow 0$ if the interaction strength exceeds a critical value. In contrast, the many-particle Landau-Zener tunneling probability always tends to zero in the adiabatic limit. However, this convergence is extremely slow so that  the breakdown of adiabaticity is approximately present also in the many-particle description. 
To highlight the origin of this breakdown we analyze the full quantum state during 
a Landau-Zener sweep in more detail  in section \ref{sec-phase} using the $SU(2)$-phase space techniques derived in \cite{07phase,07phaseappl}.  We show that many features of 
the many-particle dynamics can be captured to astonishing accuracy within the phase 
space description, including the depletion of the condensate mode as well as number 
squeezing of the final state. Yet, we show that using a  Landau-Zener sweep to generate
squeezed states for quantum metrology is very difficult for realistic systems.
  Section \ref{sec-limit}  then gives a detailed analysis of the 
region of validity of the mean-field approximation and the convergence to
the mean-field limit. In section \ref{sec-noise} we will briefly discuss the 
influence of phase noise, which is unavoidable in every experimental realization.
We conclude with a short summary and outlook.

\section{Mean-field and many-particle description of a two-mode BEC}
\label{sec-mpmf}

The Bose-Hubbard type hamiltonian 
\begin{eqnarray}
\hat H = \epsilon (t) \left( \hat a^\dagger_2 \hat a_2- \hat a^\dagger_1 \hat a_1\right) - J \left( \hat a^\dagger_1 \hat a_2 + \hat a^\dagger_2 \hat a_1\right) + \frac{U} {2} \left( \hat n_1 (\hat n_1-1)+ \hat n_2( \hat n_2-1)\right) \label{eqn-hami-bh2}
\end{eqnarray}
describes the dynamics of ultracold atoms in a double-well potential or the dynamics of a system of two-level atoms, respectively \cite{Albi05,Schu05,Milb97,Smer97}.
The operators $\hat a_1$ and $\hat a_2$ annihilate an atom in the first and second mode, respectively, while the operators $\hat n_i=\hat a^\dagger_i \hat a_i$ describe the population of the wells $i=1,2$. The tunneling matrix element and the on-site interaction strength are denoted by $J$ and $U$ and the time-dependent energy offset of the two modes is given by $2\epsilon(t) = 2 \alpha t$. In all numerical examples we shall set $\hbar=J=1$, thus measuring time in units of the tunneling time $\hbar/J$.

The time evolution generated by the hamiltonian (\ref{eqn-hami-bh2}) preserves the total particle number. It can be rewritten using the generalized angular momentum operators
\begin{eqnarray} 
  \hat L_x &=& \frac{1}{2} 
      \left( \hat a_1^\dagger \hat a_2  + \hat a_2^\dagger \hat a_1 \right), \nn \\
  \hat L_y &=& \frac{i}{2} 
      \left( \hat a_1^\dagger \hat a_2  - \hat a_2^\dagger \hat a_1 \right),  \label{eqn-angular-op}\\
  \hat L_z &=& \frac{1}{2} 
      \left( \hat a_2^\dagger \hat a_2  - \hat a_1^\dagger \hat a_1 \right), \nn   
\end{eqnarray}
which correspond to the tunneling, the momentum and the population imbalance, respectively.
In this this representation, the hamiltonian (\ref{eqn-hami-bh2}) is given by
\be
  \hat H = 2 \epsilon(t) \hat L_z - 2J \hat L_x + U \hat L_z^2. 
\ee
Initially, the two modes are energetically well separated and the ground state of the Bose-Hubbard hamiltonian (\ref{eqn-hami-bh2}) is 
\be
  |\Psi(t \rightarrow -\infty)\rangle = (N!)^{-1/2} (\hat a_1^\dagger)^N |0\rangle,
\ee
thus we assume that initially all particles are localized in the first well, corresponding to a fully condensed state. The many-particle Landau-Zener transition probability for the population is then
given by
\be
  P_{\rm LZ}^{\rm mp} := \frac{\langle \hat n_1(t \rightarrow + \infty)\rangle}{\langle
  \hat n_1(t \rightarrow - \infty)\rangle} \, .
  \label{eqn-plz-mp-def}
\ee

In the following, the many-particle quantum state is denoted by a capital $\Psi$, 
while the lower case $\psi$ is used for the components of the mean-field state
vector. To distinguish the transition probabilities we use the superscripts mp and mf for the many-particle and mean-field quantities, respectively.

In the mean-field approximaton, the time evolution is given by the discrete Gross-Pitaevskii equation
\cite{Milb97,Smer97}.
\be
  \ri \frac{\rd}{\rd t}
  \left(\begin{array}{c} \psi_{1} \\ \psi_2 \end{array} \right)
  =    \left(\begin{array}{c c}
\alpha t + g |\psi_1|^2 & -J \\ -J & -\alpha t + g |\psi_2|^2
 \end{array} \right)
  \left(\begin{array}{c} \psi_{1} \\ \psi_2 \end{array} \right),
  \label{eqn-dnlse2}
\ee
where $g=UN$ is the macroscopic interaction strength. The mean-field approximation is valid in the limit $N\rightarrow\infty$, while $g$ is kept constant.
In close analogy to the angular momentum operators (\ref{eqn-angular-op}), we define the Bloch vector
\begin{eqnarray}
  && s_x = \frac{1}{2} \left(\psi_1^* \psi_2 + \psi_2^* \psi_1 \right), \nn \\
  && s_y = \frac{\ri}{2} \left(\psi_1^* \psi_2 - \psi_2^* \psi_1 \right), \nn \\
  && s_z = \frac{1}{2} \left(\psi_2^* \psi_2 - \psi_1^* \psi_1 \right).
  \label{eqn-gpe-bloch}
\end{eqnarray}
The dynamics of the Bloch vector $\vec s$ is restricted to the surface of the Bloch
sphere, as the norm $\vec s^2 = 1/2$ is conserved by the equations of motion in the absence of phase noise. Thus, a convenient representation of the Bloch vector is given by the polar decomposition
\be
  \vec s = \frac{1}{2} \left( \begin{array}{c}
   \sin \theta  \cos \phi \\ \sin \theta \sin \phi \\  - \cos \theta
   \end{array} \right).
  \label{eqn-s-by-angle}
\ee
In this setting, the Landau-Zener tunneling probability in the level $j=1,2$ is defined as
\be
  P_{\rm LZ}^{\rm mf} := \frac{|\psi_j(t \rightarrow + \infty)|^2}{
  |\psi_j(t \rightarrow - \infty)|^2} \, .
  \label{eqn-plz-mf-def}
\ee
Again we assume that all particles are initially localized in one of the modes, i.e.
$\psi_j(t \rightarrow - \infty) = 1$.

A significant extension of the applicability of the mean-field approximation is achieved if one considers the dynamics of quantum phase space distributions instead of single mean-field trajectories. While the common mean-field approach allows only statements about expectation values, the phase space description takes also the higher moments and their time evolution approximately into account. Here, we will only review the basic definitions. For further details and a rigourous mathematical introduction see \cite{07phase,07phaseappl} and references therein.  

The starting point is the Husimi or $Q$-Function, which is definded as the projection 
onto the set of $SU(2)$-coherent states
\begin{eqnarray}
  Q(\theta,\phi,t) = |\langle \theta,\phi | \Psi(t) \rangle|^2,
  \label{eqn-def-Q}
\end{eqnarray}
with 
\begin{eqnarray}
  \ket{\theta,\phi} = \frac{1}{\sqrt{N!}} \left( \cos(\theta/2) 
    a_1^\dagger + \sin(\theta/2) 
    e^{-i \phi} a_2^\dagger \right)^N \ket{0,0}.
\end{eqnarray}
Note that the quantum state can be uniquely reconstructed from this representation, due to the 
overcompleteness of this basis set. The exact dynamics of the Husimi function is then given by
\cite{07phase,07phaseappl}
\bea   && \frac{\partial}{\partial t}  Q(\theta,\phi) = \Bigg\{ 
        2 \epsilon(t) \frac{\partial}{\partial \phi}  
        + 2J \bigg( \sin \phi \frac{\partial}{\partial \theta}
         - \cos \phi \cot \theta \frac{\partial}{\partial \phi} \bigg) \nn \\
    && \qquad \qquad \qquad \qquad - g \cos \theta \frac{\partial}{\partial \phi}
     + \frac{g}{N} \sin \theta \frac{\partial^2}{\partial \phi \partial \theta}  \Bigg\} Q(\theta,\phi) .
      \label{eqn-eom-husimi}
\eea
It is important to note that this exact evolution equations can be written as a classical Liouvillian phase space flow plus a quantum correction term which vanishes as $1/N$. The classical part is equivalent to the discrete Gross-Pitaevskii equation (\ref{eqn-dnlse2}) in the appropriate parametrization \cite{07phaseappl}. A semiclassical approximation of the phase space flow is thus provided by a truncated phase space dynamics: The initial state is mapped to its Husimi function, which is then propagated according to a classical Liouville equation omitting the quantum corrections in equation (\ref{eqn-eom-husimi}). Equivalently we will consider an ensemble of classical phase space trajectories whose starting points are distributed according to the initial Husimi function. The truncated phase space evolution defined above clearly goes beyond the common mean-field dynamics as it enables us to approximate the dynamics of the higher moments of the quantum state.

\begin{figure}[tb]
\centering
\includegraphics[width=6cm, angle=0]{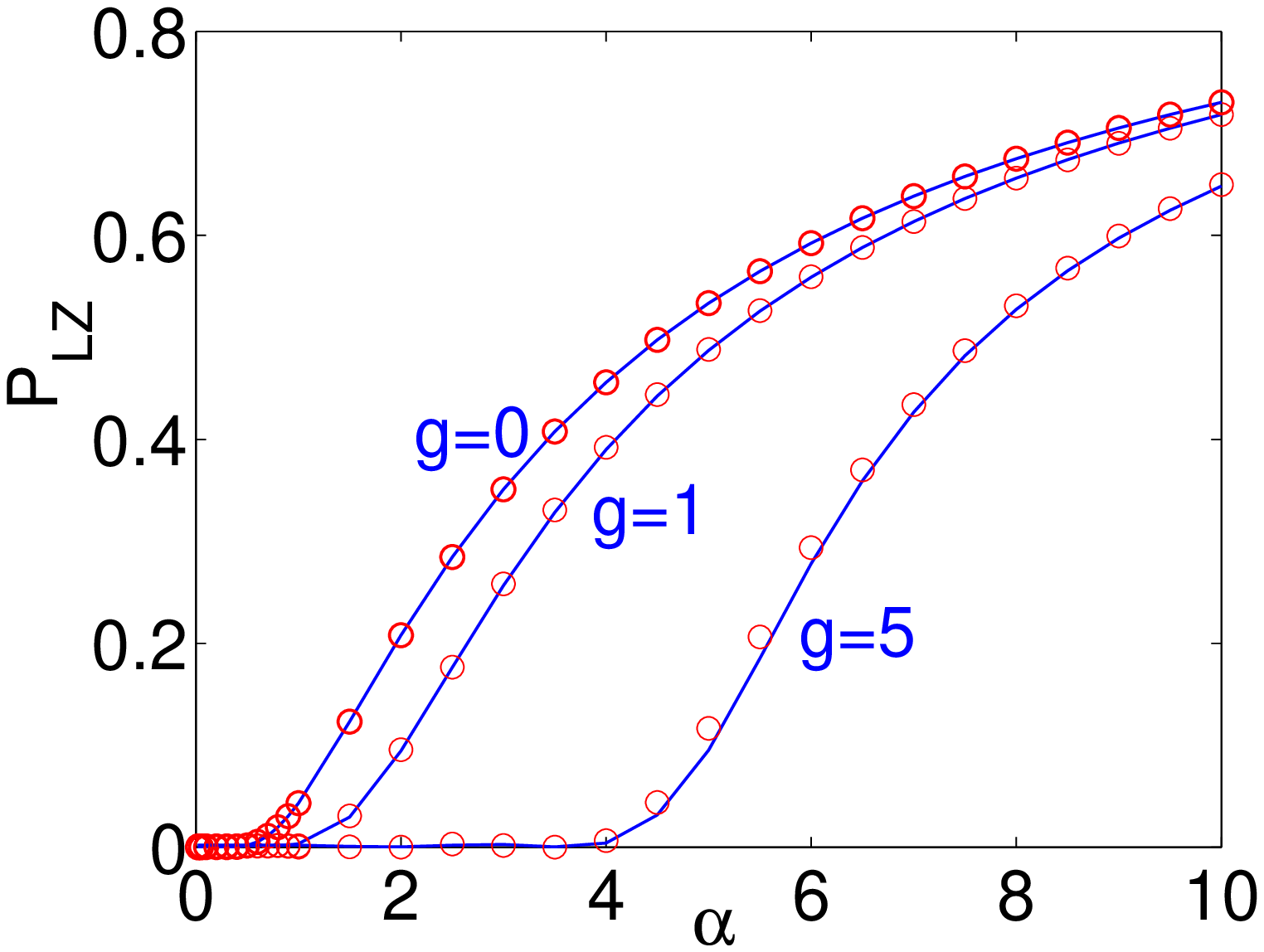}
\hspace{5mm}
\includegraphics[width=6cm, angle=0]{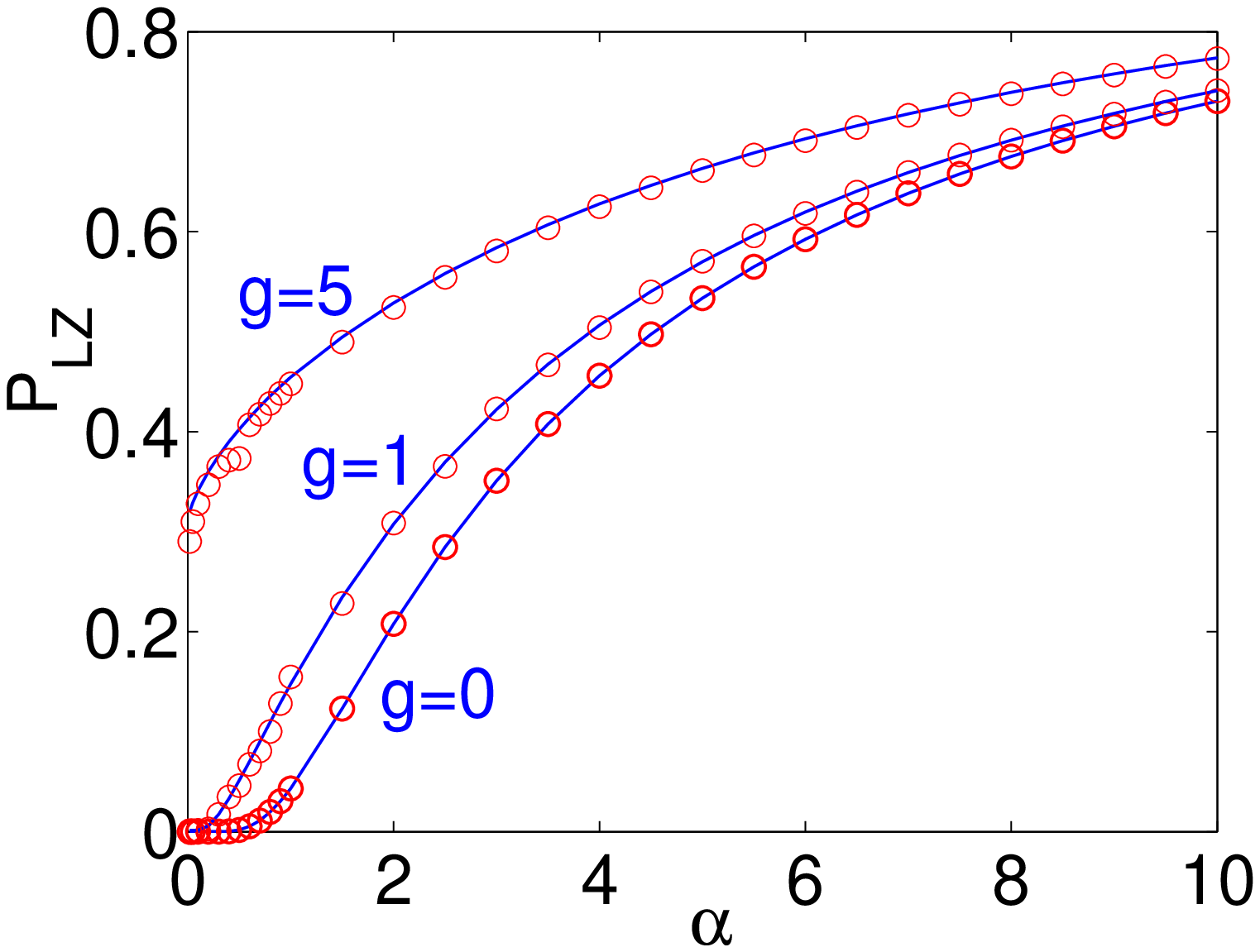}
\caption{\label{fig-plz-alpha}
Landau-Zener tunneling probability in the lower level (left) and the upper 
level (right) for $J=1$ and $g=0,1,5$. Mean-field results $P_{\rm LZ}^{\rm mf}(\alpha)$ 
are plotted as solid blue lines, many-particle results $P_{\rm LZ}^{\rm mp}(\alpha)$ 
for $N=50$ particles as red circles.}
\end{figure}

Within the phase space description, the expectation values of the generalized angular momentum operators (\ref{eqn-angular-op}) are obtained by an integration over the quasi-probability density:
\be
  \langle \hat{\vec L} \rangle = (N+2) \int \vec s(\theta,\phi) Q(\theta,\phi) \sin \theta \rd \theta \rd \phi.    
  \label{eqn-blochev-Q}
\ee
As a direct consequence, we can calculate the reduced single particle density matrix (SPDM), which is defined as 
\begin{eqnarray}
  \rho = \left( \begin{array}{c c}
  1/2 - \langle \hat L_z \rangle/N  & 
  \langle \hat L_x \rangle/N  - \ri \langle \hat L_y \rangle/N  \\
  \langle \hat L_x \rangle/N  + \ri \langle \hat L_y \rangle/N  & 
  1/2 + \langle \hat L_z \rangle/N \\
  \end{array} \right) .
\label{eqn-lz2-spdm}
\end{eqnarray}
The SPDM is a very useful quantity, since it characterizes the many-body quantum state of the trapped atoms. In particular, the fraction of atoms condensed to a single quantum state (the BEC) 
is given by the leading eigenvalue of the SPDM \cite{Legg01}.
If the the expectation value $\langle \hat{\vec L} \rangle/N$ lies on the Bloch sphere, i.e. has a magnitude of $1/2$ (as it is always the case in the common single-trajectory mean-field approach), then the two eigenvalues of the SPDM are always $\{0,1\}$ indicating a pure BEC. The phase space representation is not limited to product states. Due to the averaging procedure in equation (\ref{eqn-blochev-Q}), the expectation value of the Bloch vector is then no longer restricted to the surface, but can lie anywhere inside the Bloch sphere.
The phase space approach has been proven to be a very useful tool to go beyond the usual mean-field description \cite{07phaseappl}, especially in the description of dynamical instabilities, where it is clearly not sufficient to take into account only expectation values and to neglect all higher moments.  As we show in the following, this approach also resolves the non-commutability of the adiabatic and the semiclassical limit, which therefore must be considered as an artifact of the single-trajectory description. 

\begin{figure}[t]
\centering
\includegraphics[width=6cm, angle=0]{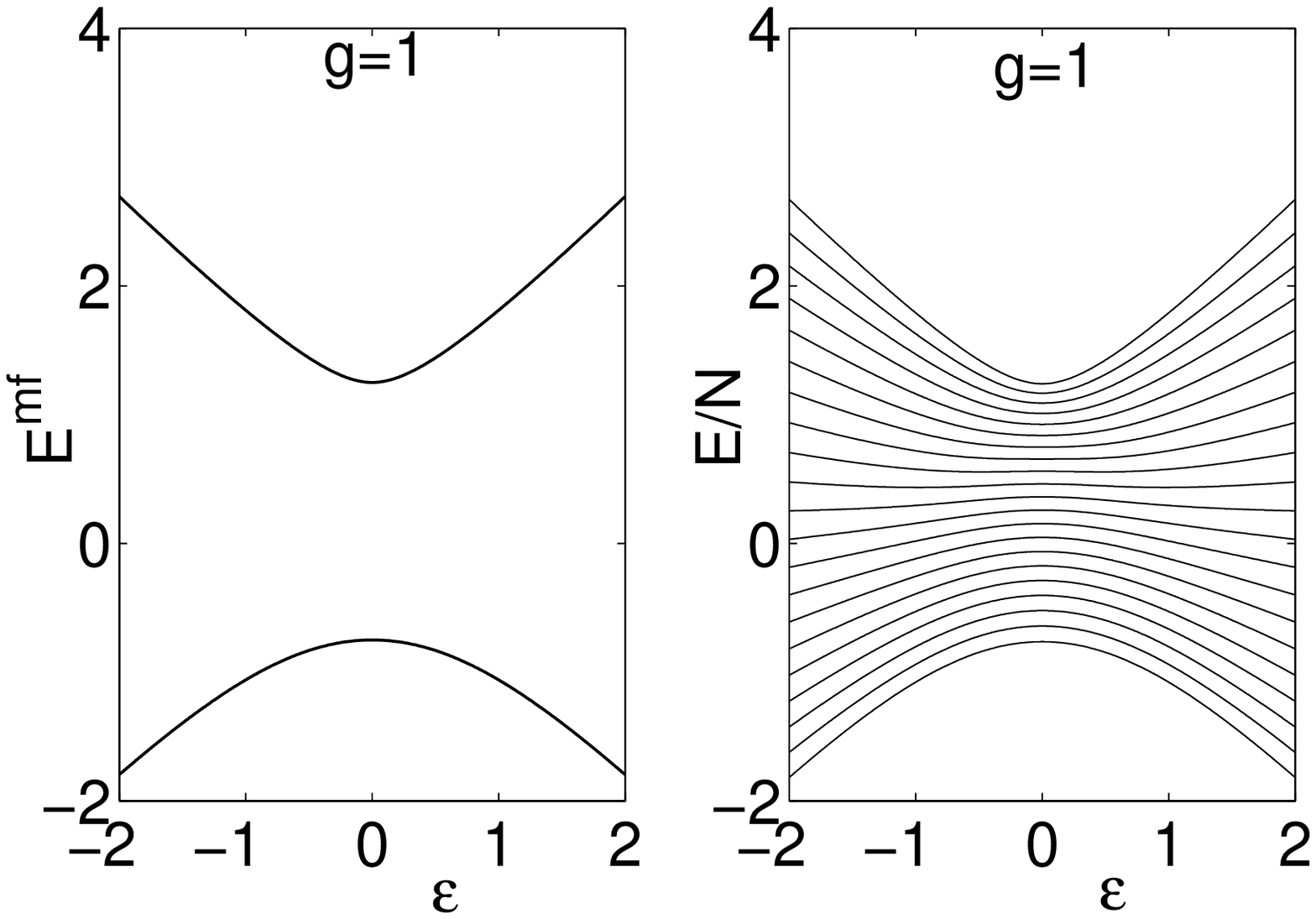}
\hspace{5mm}
\includegraphics[width=6cm, angle=0]{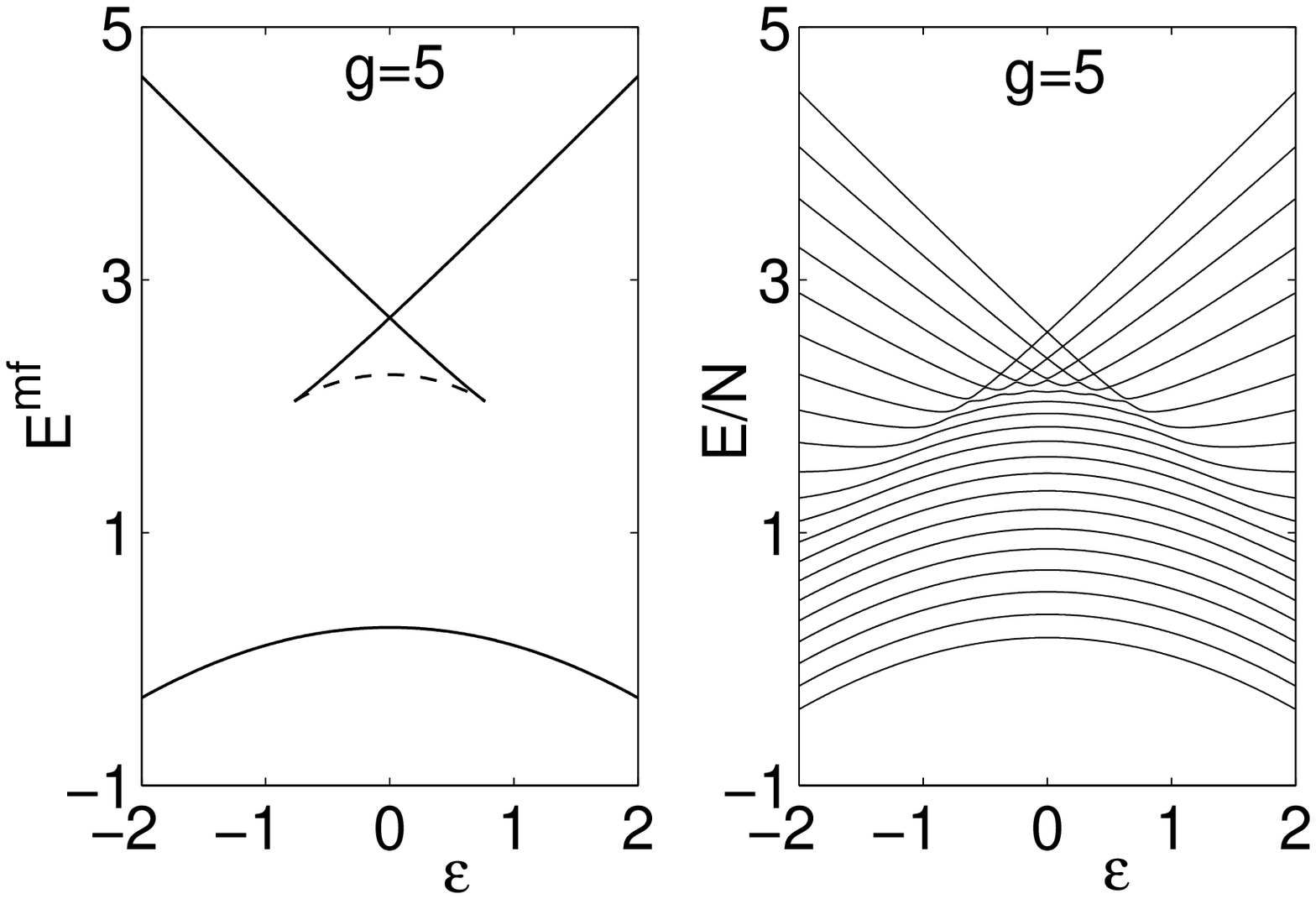}
\caption{\label{fig-levels-mpmf}
Energy of the mean-field stationary states (\ref{eigenergy_mf}) in comparison to
the eigenenergies of the two-mode Bose-Hubbard model (\ref{eqn-hami-bh2}) 
in dependence of the offset $\epsilon$ 
for $J=1$ and $g=1$ (left) and $g=5$ (right) and $N=20$ particles.}
\end{figure}

\section{Nonlinear Landau-Zener tunneling}
\label{sec-lz1}

The nonlinear self-interaction fundamentally alters the dynamics of the
system \cite{Wu00,Zoba00,Wu03} and strongly influences the Landau-Zener 
transition probability, as can be seen in figure~\ref{fig-plz-alpha}.
The solid lines show the mean-field Landau-Zener tunneling 
probability (\ref{eqn-plz-mf-def}) in dependence of the parameter 
velocity $\alpha$ for different values of the interaction strength $g$. 
For this calculation we have used the common single-trajectory mean-field 
approximation. However, there are no visible differences to the phase-space 
results for the actual parameters.
The open circles represent the corresponding many-particle results.
In the linear case $g=0$, one recovers the result (\ref{eqn-plz-lin})
for the Landau-Zener tunneling probability.
For a slow parameter variation, the state can adiabatically follow the instantaneous
eigenstates and thus most particles tunnel coherently to the other well.
For a faster sweep, this coherent tunneling effect is strongly disturbed such that 
the Landau-Zener transition probability no longer vanishes.
This effect is present in both, the transition probability in the upper and the lower level.

In the nonlinear case, the tunneling probability becomes strongly 
asymmetric: it increases as $g$ increases in the upper level, while it decreases 
in the lower level. 
To understand this effect, it is insightful to consider the total energy of the mean-field system. 
Figure \ref{fig-levels-mpmf} 
shows the eigenenergies of the Hamiltonian (\ref{eqn-hami-bh2}) in comparison 
to the total energies of the `nonlinear eigenstates', i.e.~the stationary states of 
the mean-field dynamics (\ref{eqn-dnlse2}),
\begin{eqnarray} \label{eigenergy_mf}
   E^{\rm mf}=\epsilon (|\psi_2|^2-|\psi_1|^2) - J(\psi_1^*\psi_2 + \psi_2^*\psi_1) 
    + \frac g 2 (|\psi_1|^4+|\psi_2|^4).
\end{eqnarray}
Compared to the non-interacting case $g=0$, the left-hand side shows that 
the upper level is sharpended, while the lower level is flatened for small interactions $0<g<2J$.
This flattening supresses the tunneling probability from the lower level to the upper level, leading to a
decreased Landau-Zener probability in the adiabatic regime. On the other hand, the sharpening of the upper 
level makes it more difficult to follow the adiabatic eigenstates, which results in a increased 
Landau-Zener probability for the upper level, as can be seen on the right-hand side of figure~\ref{fig-plz-alpha}.

Most remarkably, the tunneling probability in the 
upper level does not even vanish in the adiabatic limit 
$\alpha \rightarrow 0$ for $g>2J$, i.e. adiabaticity breaks down in the strongly interacting case.

\begin{figure}[t]
\centering
\includegraphics[width=12cm, angle=0]{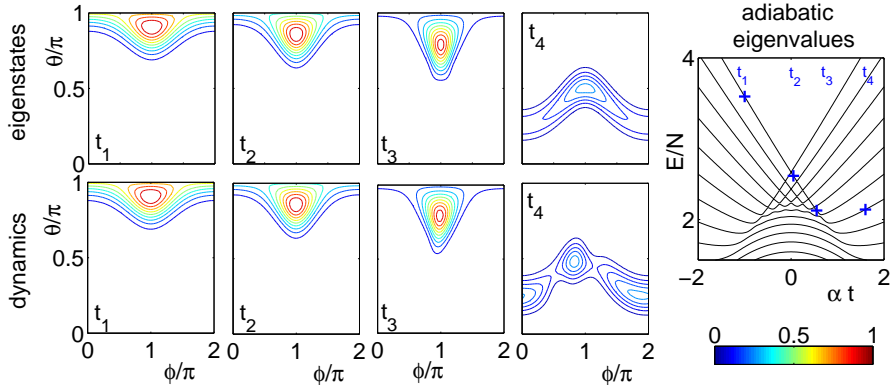}
\caption{\label{fig-lz-hus-dyn}
Dynamics of the many-particle Landau-Zener transition for $J=1$, $g=5$, $N=20$
and $\alpha = 0.1$.
The upper figures show a contour plot of the Husimi distribution of the 
instantaneous eigenstates marked in the level scheme on the right at 
times $t_j = -10, 0.5, 5.5, 16$.
The lower figures show the Husimi distribution of the dynamical quantum state 
$|\Psi(t)\rangle$ at the same times.}
\end{figure}

In order to explore the origin of this breakdown of adiabaticity we compare again
the eigenstates of the many-particle system to 
the stationary states of the mean-field system. 
For $g>2J$ the mean-field eigenenergies show a swallow's tail structure in the upper level, 
reflecting the occurrence of a bifurcation of one of the steady states
into three new ones, one of them hyperbolically unstable (dashed line) and two 
elliptically stable (solid lines). The system can adiabatically follow the steady states 
as long as these are elliptically stable. This is possible only until the end of the 
swallow's tail where the elliptic fixed point vanishes in an inverse bifurcation with 
the hyperbolic fixed point \cite{Zoba00,Liu02}. Then the dynamics becomes 
unstable and adiabaticity is lost even for very small values of $\alpha$.

The swallow's tail in the mean-field energy corresponds to a caustic of the many-particle 
eigenenergy curves in the limit $N \rightarrow \infty$, which are bounded by the 
mean-field energies from below and above. 
Within this caustic one finds a series of quasi-degenerate avoided crossings of the 
many-particle levels. The level splitting at these crossings tends to zero exponentially
fast in the mean-field limit $N \rightarrow \infty$ with $g = UN$ fixed 
\cite{06zener_bec,Wu06}. Thus the system 
will show a complete diabatic time evolution at these quasi-crossings even for very 
small values of $\alpha$. Outside the swallow's tail one finds common avoided 
crossings, where the system evolves adiabatically for small value of $\alpha$.

Note, however, that the breakdown of adiabaticity  is only approximate for the 
many-particle system. It is known that for a symmetric tridiagonal Hamiltonian, such as 
the one we are considering (\ref{eqn-hami-bh2})  with $J \ne 0$,  the level spacings 
in the spectrum may be exponentially small but nevertheless always non-zero \cite{Wilk65}. 
Thus adiabaticity can be restored when the 
parameter velocity $\alpha$ is decreased well below the square of the 
residual level splitting $\Delta$ \cite{Wu06}:
\be
  \alpha \stackrel{!}{\ll} \Delta^2 \qquad \mbox{with} \qquad 
  \Delta \propto N \exp(-\eta N),
\ee
where $\eta$ is a proportionality constant which depends of the system parameters.
However, the adiabaticity condition on the velocity becomes exponentially
difficult to fulfill.
Thus the breakdown of adiabaticity is also present in the full many-particle 
system for any realistic set of parameters.
The same dynamics is found for attractive nonlinearities, $g<0$, only the
roles of the upper and lower level are exchanged.

\section{Phase space picture}
\label{sec-phase}

\begin{figure}[t]
\centering
\includegraphics[width=6cm, angle=0]{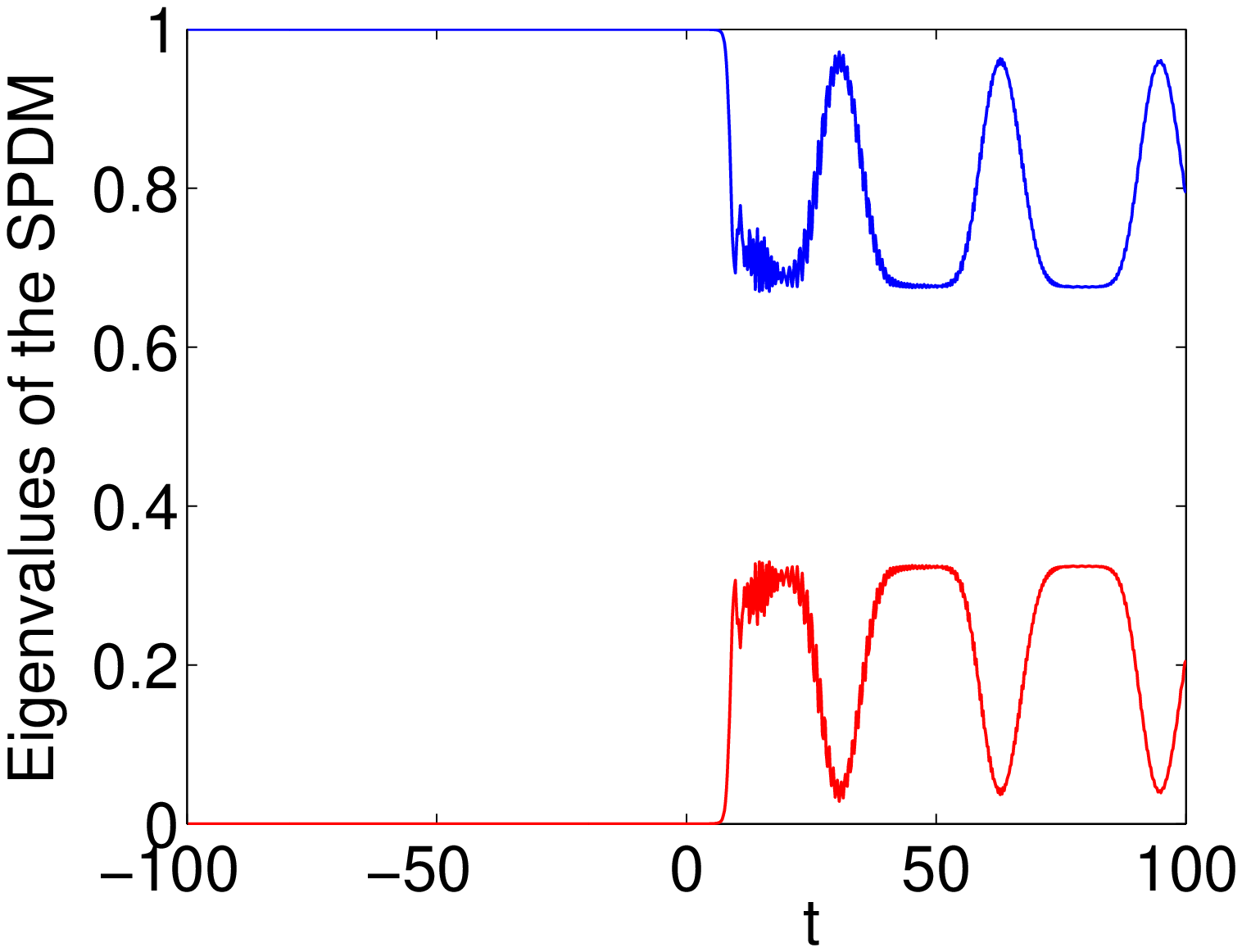}
\hspace{5mm}
\includegraphics[width=6cm, angle=0]{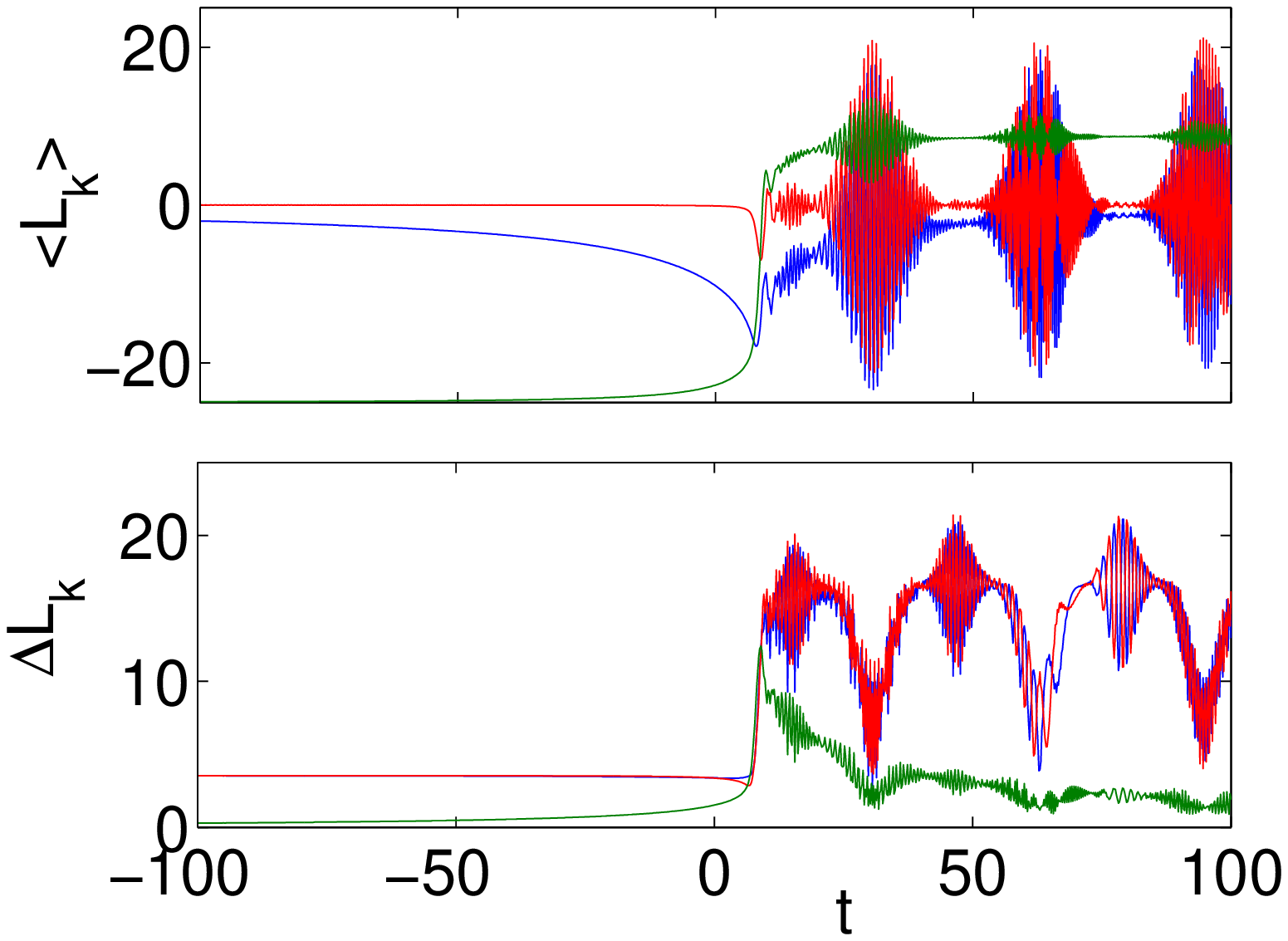}
\caption{\label{fig-lz-dyn-mp}
Dynamics of the many-particle Landau-Zener transition for $g=5$, $N=50$
and $\alpha = 0.1$.
The left figure shows the eigenvalues of the SPDM (\ref{eqn-lz2-spdm}).
The expectation values $\langle \hat L_k \rangle$ and widths $\Delta \hat L_k$ 
of the angular momentum operators (\ref{eqn-angular-op}) for $k=x$ (blue), 
$k=y$ (red) and $k=z$ (green) are plotted on the right-hand side.
The onset of the oscillatory dynamics corresponds to the cusp of the swallow-tail 
structure in the mean-field representation (cf. figure \ref{fig-levels-mpmf}).}
\end{figure}

Further insight into the dynamics of nonlinear Landau-Zener tunneling can 
be gained within the phase space picture introduced in section \ref{sec-mpmf} 
and in \cite{07phase,07phaseappl}.

According to the remarks in the previous section, the system will undergo a series of 
diabatic transitions up to the end of the swallow's tail and evolve adiabatically 
afterwards. To verify these claims, we compare the actual many-particle quantum 
state $|\Psi(t)\rangle$ to the instantaneous eigenstate in figure \ref{fig-lz-hus-dyn} 
at four points in time during a Landau-Zener passage. To visualize the quantum 
states, we use the Husimi distribution $Q(\theta,\phi,t)$ as defined in equation 
(\ref{eqn-def-Q}). The  right-hand side of figure \ref{fig-lz-hus-dyn} illustrates the 
series of diabatic/adiabatic transitions and the specific instantaneous 
eigenstates shown in upper panels of the figure.
One observes a good agreement between the dynamical state 
and the instantaneous eigenstates, during the transition as well as afterwards.
However, the crossover from diabatic to adiabatic transitions is not 
absolutely sharp. The final state contains small contributions from
other instantaneous eigenstates.

\begin{figure}[t]
\centering
\includegraphics[width=6cm, angle=0]{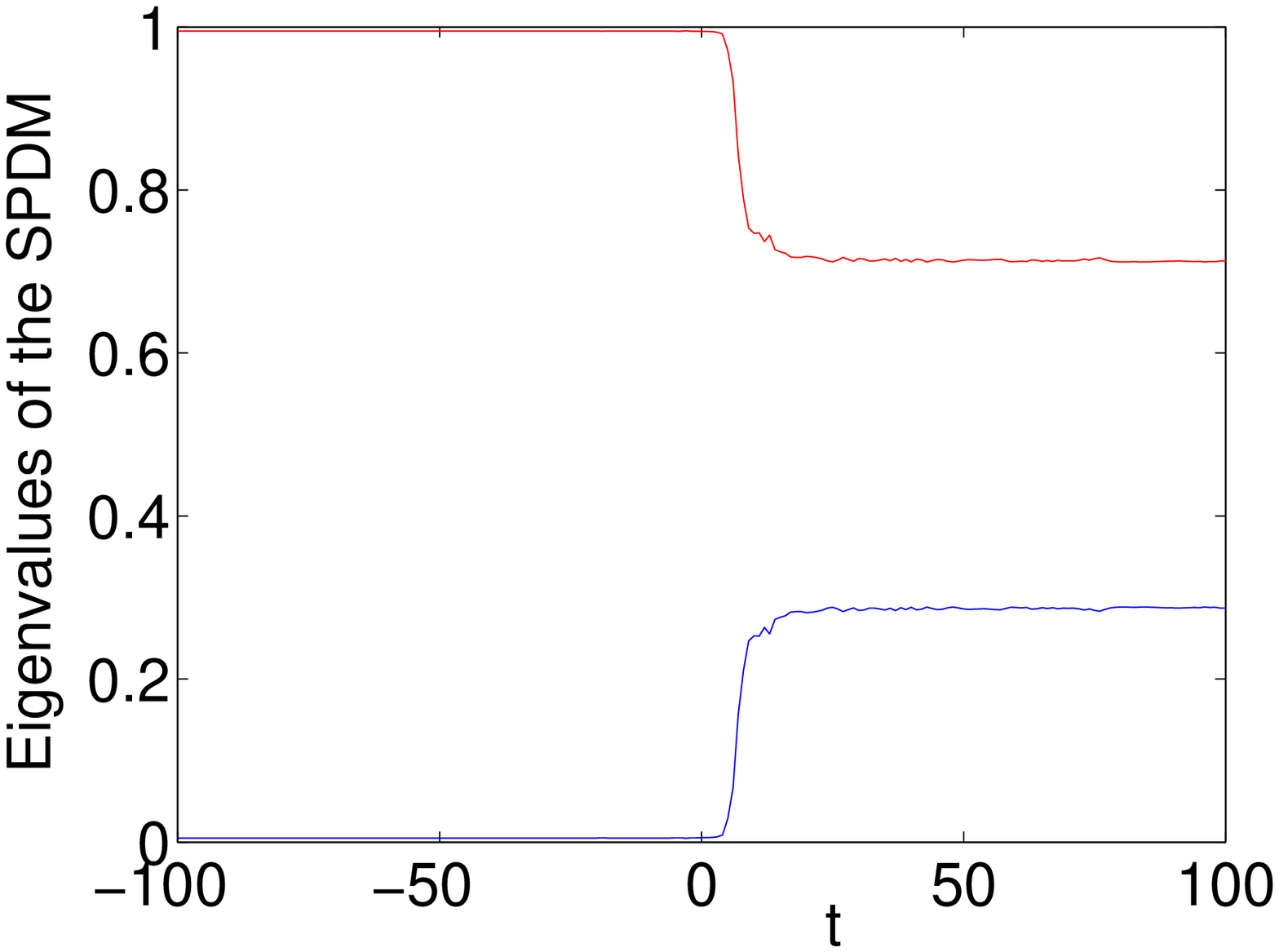}
\hspace{5mm}
\includegraphics[width=6cm, angle=0]{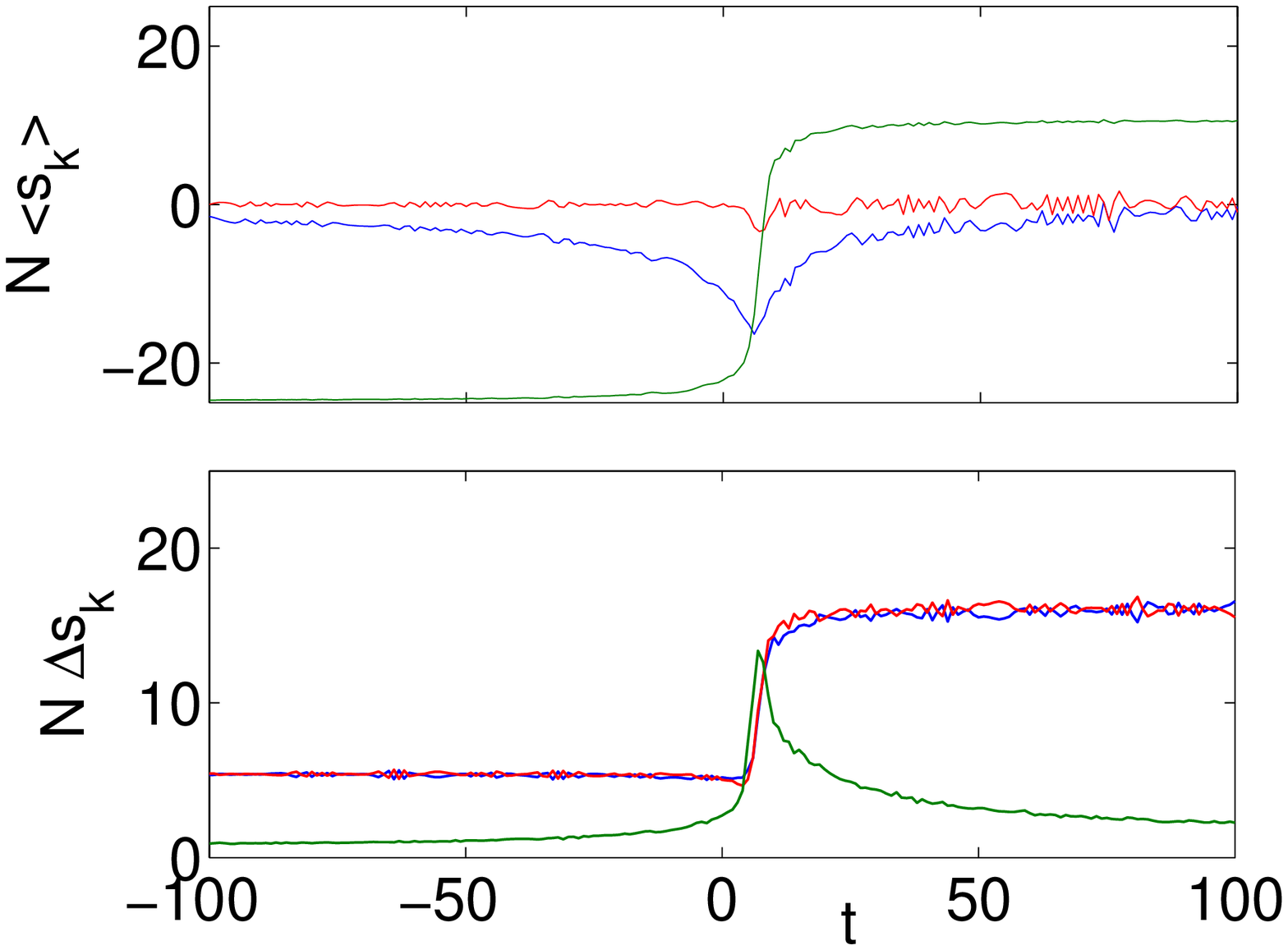}
\caption{\label{fig-lz-ensdyn}
Semiclassical simulation of the many-particle Landau-Zener transition for 
$J=1$, $g=5$, $N=50$ and $\alpha = 0.1$. 
The left figure shows the eigenvalues of the ensemble approximation for the
SPDM $\rho_{kl} = \langle \psi_k^* \psi_l \rangle_{\rm cl}$.
The ensemble expectation values $N \langle s_k \rangle_{\rm cl}$ and the
standard deviation $N \Delta s_k$ of the Bloch vector (\ref{eqn-gpe-bloch})
for $k=x$ (blue), $k=y$ (red) and $k=z$ (green) are plotted on the 
right-hand side.}
\end{figure}

In order to characterize the many-particle quantum state during the Landau-Zener 
transition, we have plotted the eigenvalues of the SPDM (\ref{eqn-lz2-spdm}) on
the left-hand side of figure \ref{fig-lz-dyn-mp}. One eigenvalue remains equal to 
unity, while the other one vanishes, indicating a fully coherent state until the 
crossover from diabatic to adiabatic transitions. Then one observes an oscillation
of the SPDM eigenvalues: The contributions of the different many-particle 
eigenstates de- and rephase periodically giving rise to a beat signal which is 
genuinely quantum. The oscillation of the coherence 
is mirrored in the evolution of the uncertainties of the angular momentum 
operators $\Delta \hat L_x$  and $\Delta \hat L_y$ shown on the right-hand side of 
figure \ref{fig-lz-dyn-mp}. The uncertainties are 
strongly enhanced when the coherence is (partly) lost. This behaviour 
can be intuitively explained in terms of the dynamics of the Husimi distribution.
The centre of mass of the Husimi function oscillates rapidly in the 
$\phi$-direction, leading to oscillations of the expectation 
values $\langle \hat L_x \rangle$ and $\langle \hat L_y \rangle$.
Furthermore the distribution breathes in the $\phi$-direction 
at a slower timescale, leading to the oscillations of the width 
$\Delta \hat L_x$ and $\Delta \hat L_y$  and the periodic revivals 
of the coherence. The oscillations of the expectation values die out 
at the times when the Husimi function is spread nearly uniformly 
in the $\phi$-direction, i.e. at the times where the coherence is minimal.
In contrast, the Husimi distribution is well localized in the $\theta$-direction 
for long times and the corresponding uncertainty $\Delta \hat L_z$ remains small. 
The population difference $\langle \hat L_z \rangle$ is thus well described
by the simple Bogoliubov mean-field approximation.
Many-particle and mean-field results for the Landau-Zener tunneling rate 
show an excellent agreement (cf. figure \ref{fig-plz-alpha}), because they 
depend only on the population difference and not on the coherence.

\begin{figure}[t]
\centering
\includegraphics[width=12cm,  angle=0]{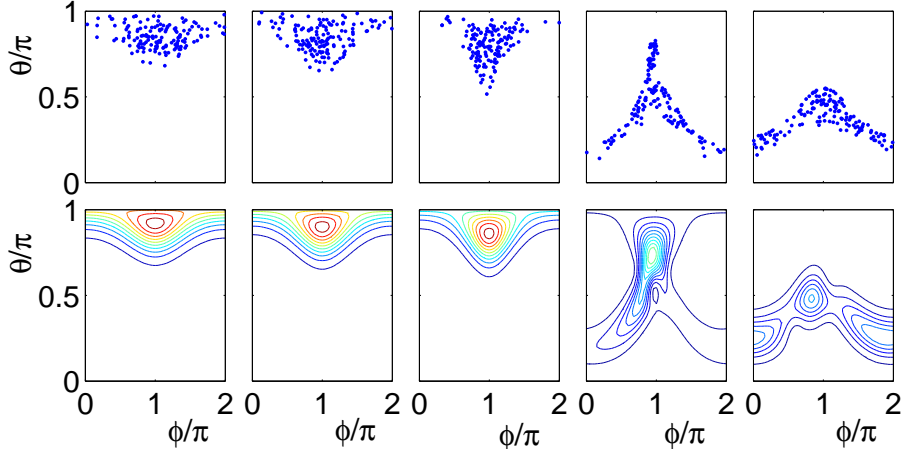}
\caption{\label{fig-lz-cphase}
The many-particle Landau-Zener scenario in phase space. The dynamics 
of an ensemble of 150 classical trajectories (upper panels) is compared 
to the evolution of the Husimi distribution (lower panels) at times 
$t=-16,-8,0,8,16$ (from left to right). Parameters are chosen as in 
figure \ref{fig-lz-hus-dyn}.}
\end{figure}

The evolution of the coherence and the uncertainties $\Delta \hat L_x$ and 
$\Delta \hat L_y$ certainly goes beyond the Bogoliubov mean-field approximation,
but most of the effects can be taken into account by the semi-classical phase space 
approach introduced in section \ref{sec-mpmf} and in \cite{07phase,07phaseappl}. 
Figure \ref{fig-lz-cphase} shows the dynamics of the many-particle Landau-Zener 
scenario in quantum phase space in comparison to the dynamics of a classical 
phase space ensemble. The expectation values and variances of the Bloch
vector $\hat{\vec L}$ calculated from such an ensemble simulation are plotted
in figure \ref{fig-lz-ensdyn}.
It is observed, that the spreading of the Husimi distribution in the direction of 
the relative phase $\phi$ and the loss of coherence are well reproduced by 
the classical ensemble.  However, the quantum beat oscillations of the coherence 
are of course not present in the classical distributions as shown in figure 
\ref{fig-lz-ensdyn}. 
The expectation value and the fluctuations of the classical Bloch vector 
$\vec s$ defined in equation (\ref{eqn-gpe-bloch}) show a similar effect.
The global dynamics of the angular momentum operator $\hat{\vec L}$ plotted
in figure \ref{fig-lz-dyn-mp} is well reproduced, whereas all the
quantum beats are absent. These are genuine many-particle 
quantum effects.

\begin{figure}[t]
\centering
\includegraphics[width=12cm, angle=0]{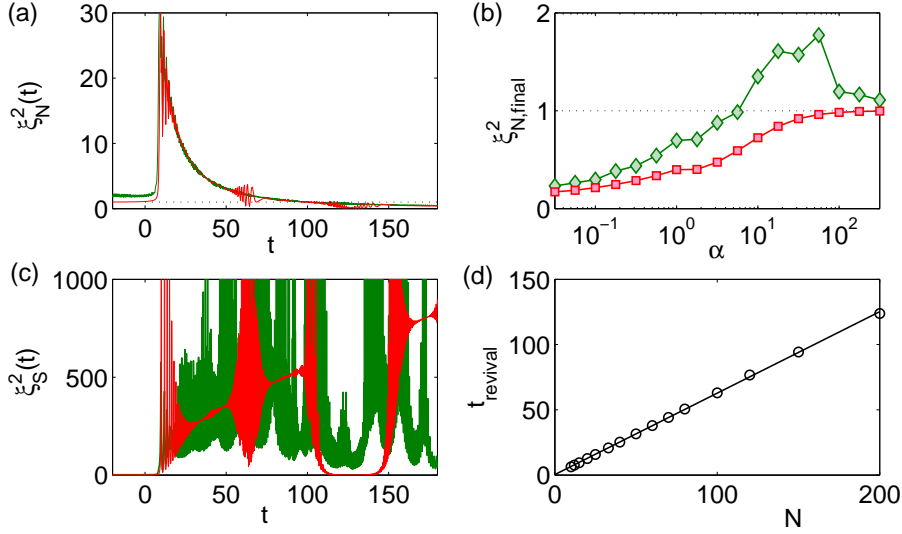}
\caption{\label{fig-squeezing}
Number squeezing during a Landau-Zener sweep. Many-particle results 
(red) are compared to an ensemble simulation (green).
(a) Evolution of the relative squeezing parameter $\xi_N^2$ for 
a slow sweep with $\alpha = 0.1$.
(b) Final value of $\xi_N^2$  after the sweep as a function of
the parameter velocity $\alpha$.
(c) Evolution of the spectroscopic squeezing parameter $\xi_S^2$ 
for a slow sweep with $\alpha = 0.1$.
(d) Dependence of the revival time of the occurrence  of a squeezed state on the particle number $N$.
The remaining parameters are $J=1$, $g=-5$ and $N=200$ particles. 
}
\end{figure}

The previous results show that the many-particle quantum state after 
a nonlinear Landau-Zener sweep is far from being a pure BEC. 
In particular it has been claimed that the final state is strongly
number squeezed in comparison to a pure BEC with the same
density distribution \cite{Smit09}. The figures \ref{fig-lz-dyn-mp} 
and \ref{fig-lz-ensdyn} show the evolution of the expectation 
values and variances of $\hat{\vec L}$, comparing many-particle 
results to a phase space approximation. One observes that the
number fluctuations $\Delta L^2_{z}$ are strongly increased during 
the sweep, but relax to a smaller value again afterwards. This 
evolution is well described within the semiclassical phase space 
picture.
A further quantitative analysis of number squeezing during a 
Landau-Zener sweep is provided in figure \ref{fig-squeezing}, 
comparing exact results (red) to an ensemble simulation (green).
For a pure BEC with a given particle density, number fluctuations 
are given by  $\Delta L^2_{z,\rm ref} = \langle \hat n_1 \rangle 
\langle \hat n_2 \rangle /N$. Thus one can define the parameter 
$\xi_N^2 = \Delta L^2_{z}  /  \Delta L^2_{z,\rm ref}$,
which measures the suppression of number fluctuations 
in comparison to a pure BEC.
Figure \ref{fig-squeezing} (a) shows the value of $\xi_N^2$ 
during a slow Landau-Zener sweep with $\alpha = 0.1$. 
Indeed, $\xi_N^2$ drops well below one for long times 
indicating number squeezing. Again, this feature is well
reproduced by a phase space simulation (green).
The final value of $\xi_N^2$ after the sweep is shown in
Figure \ref{fig-squeezing} (b) as a function of the parameter 
velocity $\alpha$. Number squeezing with $\xi_N^2 < 1$ is 
observed for small values of $\alpha$ in the regime of the 
breakdown of adiabaticity, e.g. for large interaction strength, $2J<g$ . 
The phase space simulation
overestimates the variances and thus also $\xi_N^2$, but gives 
the correct overall behaviour.
For fast sweep, $\xi_N^2$ tends to one as the state remains 
approximately coherent.

However, an application of number squeezing in quantum metrology
requires a reduction of number fluctuations as well as a large phase 
coherence. Thus, a quantum state is defined to be spectroscopically 
squeezed if and only if
\be
  \xi_S^2 := N \frac{\Delta \hat L_z^2}{
    \langle \hat L_x\rangle^2  +  \langle \hat L_y\rangle^2}
     < 1.
\ee
Spectroscopic squeezing indicates multipartite entanglement  
of the trapped atoms \cite{Sore01,Este08}.
The evolution of the squeezing parameter $\xi_S^2$ during 
a slow Landau-Zener sweep with $\alpha = 0.1$ is plotted 
in figure \ref{fig-squeezing} (c). 
While the number fluctuations $\Delta \hat L_z^2$ assume 
a small constant value after the sweep, the phase coherence 
$\langle \hat L_x\rangle^2  +  \langle \hat L_y\rangle^2$ strongly 
oscillates due to the periodic de- and rephasing of the 
many-particle eigenstates (cf. figure \ref{fig-lz-dyn-mp}). 
True spin squeezing with $\xi_S^2 < 1$ is present only 
temporarily in the periods of maximum phase coherence.
The timescale of the occurence of these minima depends 
linearly on the particle number $N$, as shown in figure 
\ref{fig-squeezing} (d). For macroscopic particle numbers it takes very long
before the states rephase such that $\xi_S^2<1$ is observed. 
Moreover these revivals are extremely sensitive to phase
noise. Thus, it is doubtful that for realistic particle numbers 
Landau-Zener sweeps may be useful to generate squeezed 
states in a controlled way. 
Finally we note that the revivals of the phase coherence 
are not described by the phase space picture. Even small fluctuations 
in the phase coherence lead to large errors. Therefore the phase 
space approximation cannot account for the
short periods where true spin squeezing $\xi_S^2 < 1$ 
is observed.

Let us finally investigate the global dependence of the Landau-Zener 
tunneling rate on the interaction strength $g = UN$ in more detail. 
To this end we calculate the quantum and the classical tunneling rates given 
by equations (\ref{eqn-plz-mp-def}) and (\ref{eqn-plz-mf-def}), respectively, 
as well as the eigenvalues of the SPDM (\ref{eqn-lz2-spdm}). 
We consider an initial state that is localized in the upper level for 
$t\rightarrow - \infty$ so that adiabaticity breaks down for a repulsive 
nonlinearity $g > 2J$. 
As discussed above, a change of the sign of the interaction strength $g$ 
corresponds to an interchange 
of the two modes. For an attractive nonlinearity, adiabaticity breaks down 
in the lower level instead.
Thus we obtain a global picture of the dynamics either by calculating the 
tunneling rate in the upper and the lower level for $g>0$, or by calculating
the tunneling rate in the upper level alone for $g>0$ and $g<0$. In the 
following we choose the latter option.

\begin{figure}[t]
\centering
\includegraphics[width=6cm, angle=0]{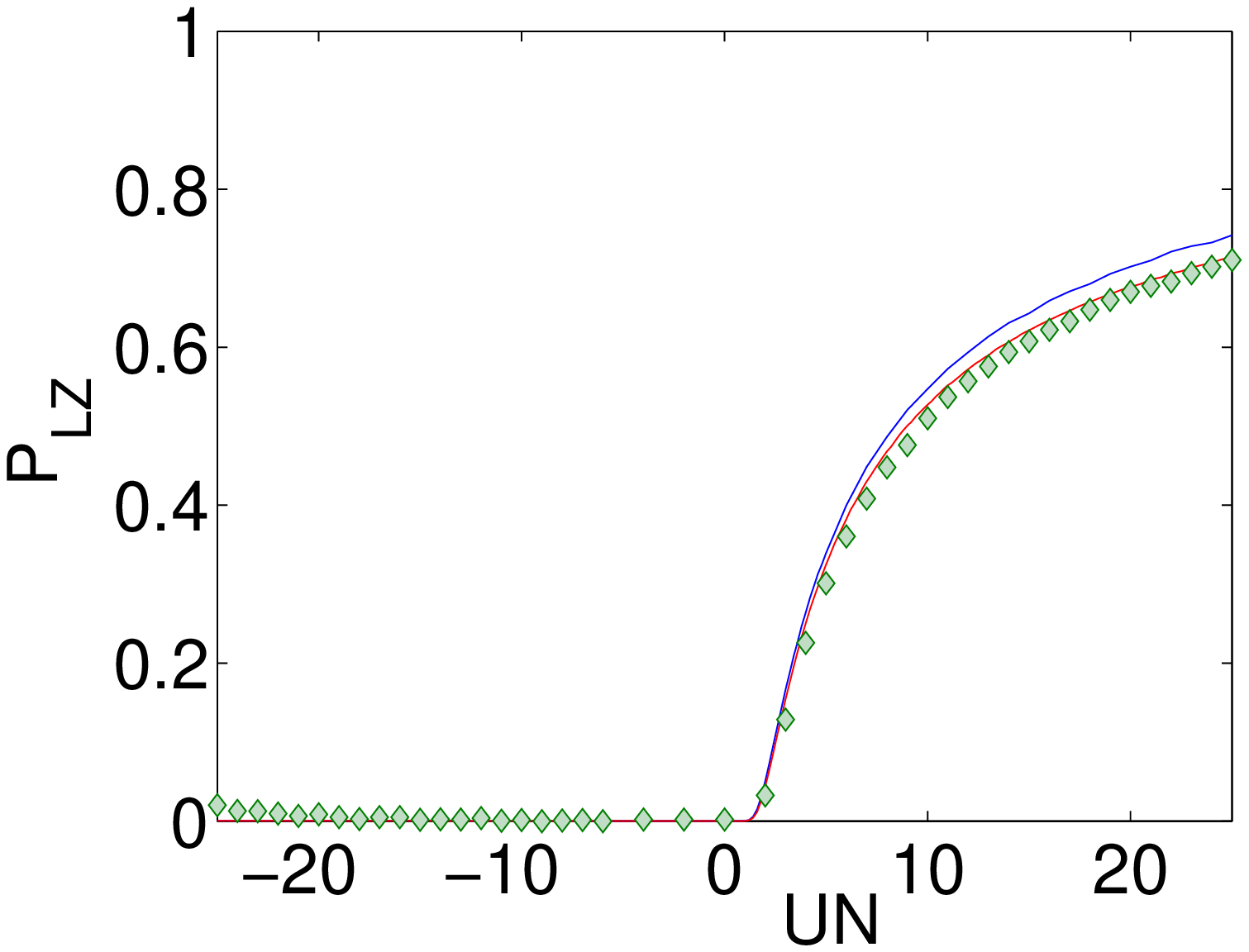}
\hspace{5mm}
\includegraphics[width=6cm, angle=0]{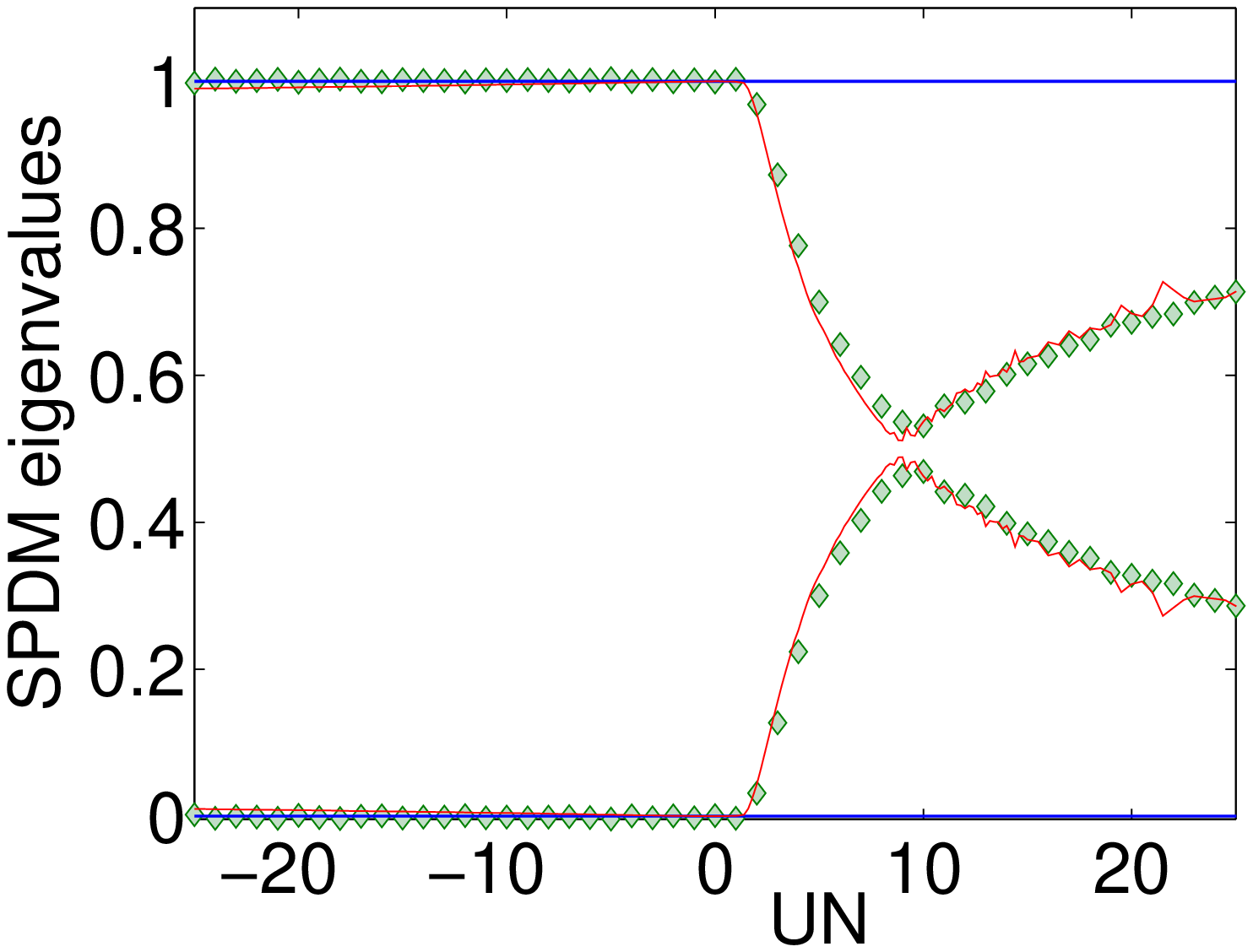}
\caption{\label{fig-plz-g-a01}
Left panel: Landau-Zener tunneling probability as a function of of the interaction strength
$UN$ for a slow parameter variation $\alpha = 0.1$ and $J=1$, $N=50$.
Right panel: Eigenvalues of the SPDM  for $t \rightarrow +\infty$.
The exact many-particle results (red line) are compared to a
phase space ensemble simulation (green diamonds) and
the single-trajectory mean-field results (blue line). }
\end{figure}

Figure \ref{fig-plz-g-a01} shows the results for $J=1$ and $\alpha = 0.1$, where 
the linear system evolves completely adiabatically. The left-hand side shows the 
many-particle and mean-field Landau-Zener tunneling probabilities as defined in 
equation (\ref{eqn-plz-mp-def}) and (\ref{eqn-plz-mf-def}), respectively.
The right-hand side shows the eigenvalues of the SPDM (\ref{eqn-lz2-spdm})
for $t \rightarrow + \infty$. Note, however, that the eigenvalues of the SPDM 
oscillate for $t >0 $ as shown in figure \ref{fig-lz-dyn-mp}, indicating a periodic
loss and revival of coherence. Figure \ref{fig-plz-g-a01} shows the 
eigenvalues of the SPDM for large times, omitting the temporal revivals explicitly.
As expected, adiabaticity breaks down as soon as $g > 2J$ and the Landau-Zener
tunneling rate increases with $g$. In the adiabatic regime, one eigenvalue of 
the SPDM is close to zero, indicating a fully coherent state. Coherence is 
lost when the adiabaticity breaks down and particles are scattered out of the condensate mode.

\begin{figure}[t]
\centering
\includegraphics[width=6cm, angle=0]{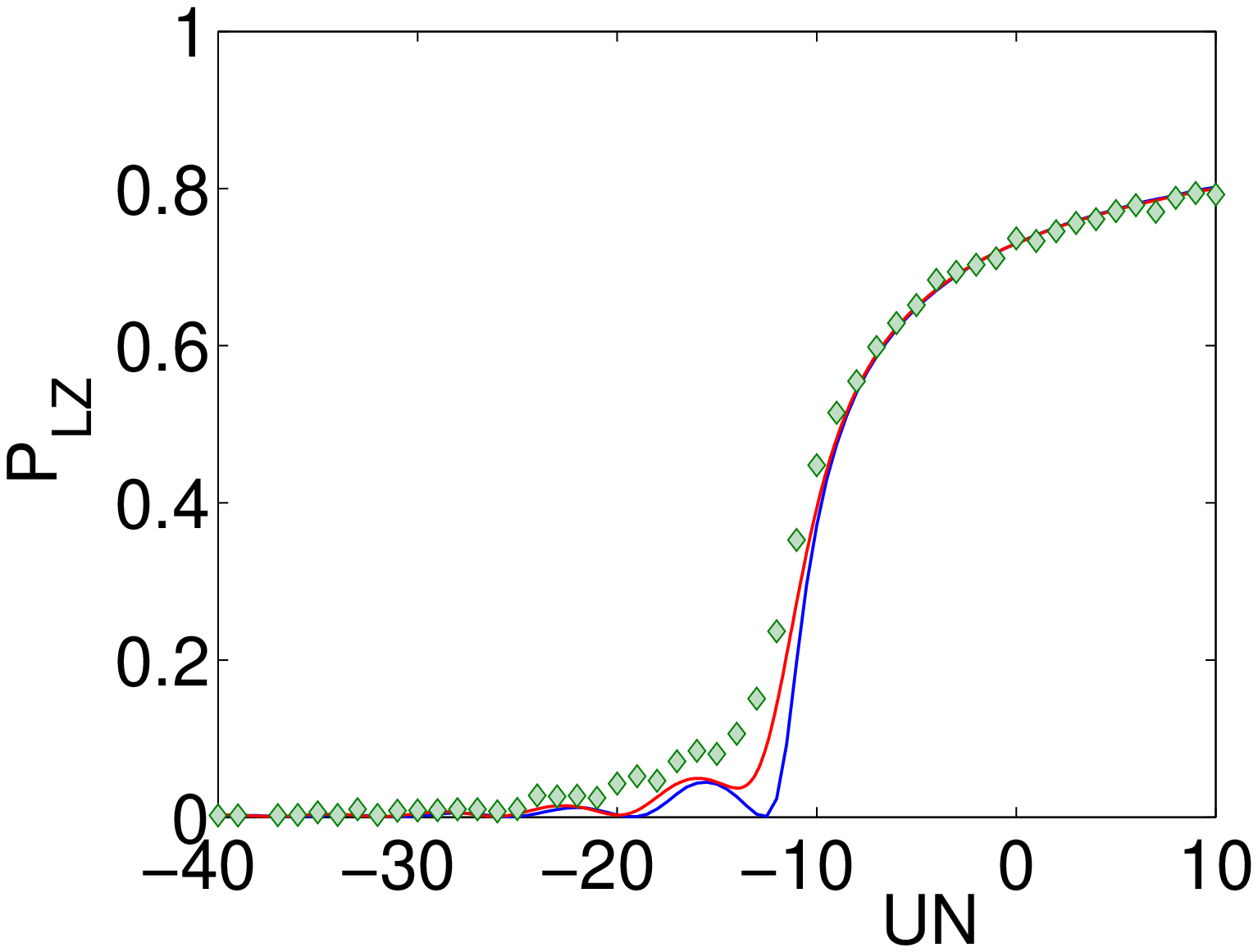}
\hspace{5mm}
\includegraphics[width=6cm, angle=0]{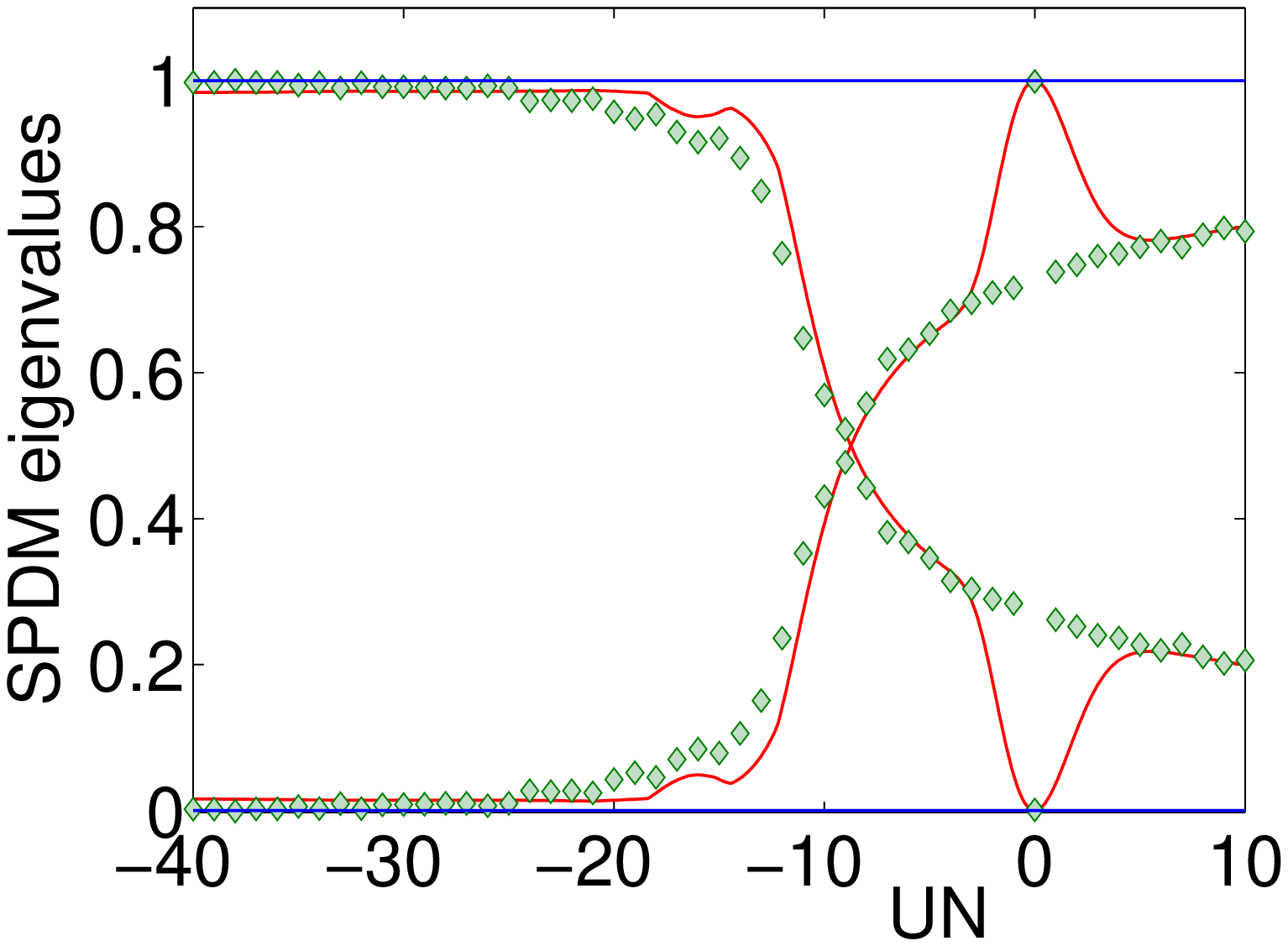}
\caption{\label{fig-plz-g-a10}
Left panel: Landau-Zener tunneling probability as a function of of the interaction strength
$g=UN$ for a fast parameter sweep $\alpha = 10$ and $J=1$, $N=50$.
Right panel: Eigenvalues of the SPDM  for $t \rightarrow +\infty$.
The exact many-particle results (red line) are compared to a
phase space ensemble simulation (green diamonds) and
the single-trajectory mean-field results (blue line). }
\end{figure}

Figure \ref{fig-plz-g-a10} shows the results for a fast sweep $\alpha = 10$ for 
$J=1$. In the linear case, equation (\ref{eqn-plz-lin}) predicts a Landau-Zener tunneling 
rate of $P_{\rm LZ} =  0.7304$. Surprisingly, the basic structure of the numerical results is very similar 
to the adiabatic case shown in figure \ref{fig-plz-g-a01}. The curves are shifted, but 
the general progression remains the same.
This is understood as follows. As argued above, an attractive nonlinearity flattens 
the upper level so that Landau-Zener tunneling is decreased. The current example 
shows that this effect is so strong that the tunneling process is completely 
suppressed so that $P_{\rm LZ} \rightarrow 0$ for large negative values of $g$. 
On the contrary, a repulsive nonlinearity leads to an increase of $P_{\rm LZ}$.
The transition between an effectively adiabatic and non-adiabatic dynamics 
occurs at $g = 2J$ for a slow parameter variation $\alpha \rightarrow 0$.
For a fast sweep, $P_{\rm LZ}$ is non-zero in the linear case $g=0$.
However, a strong attractive nonlinearity can flatten the level so much that
adiabaticity is restored again. Thus one can always enforce an adiabatic 
transition, but the necessary interaction strength $|g|$ increases monotonically
with $\alpha$.
This behaviour is also reflected in the coherence properties of the final
state shown on the right-hand side of figure \ref{fig-plz-g-a01} and 
\ref{fig-plz-g-a10}.

One astonishing feature observed in the figures \ref{fig-plz-g-a01} and 
\ref{fig-plz-g-a10} is the excellent agreement of the Landau-Zener tunneling 
rate $P_{\rm LZ}$ and the eigenvalues of the SPDM. Deviations are only 
found around $g = 0$ in figure \ref{fig-plz-g-a10}. 
This can be understood by a loss of the coherence between the two modes
for long times, i.e. 
\be
  \langle \hat a_1^\dagger \hat a_2 \rangle \rightarrow 0 \qquad \mbox{for} \quad
  t \rightarrow + \infty,
\ee
if we do not take into account for the temporal revivals illustrated
in figure \ref{fig-lz-dyn-mp}. This happens either if the atoms are not 
in a coherent state any longer or if all atoms are localized in one of the modes.
In any case we can rewrite the reduced SPDM as
\bea \label{eqn_applzSPDM}
  \rho(t \rightarrow + \infty) &\approx& 
   \left( \begin{array}{c c}
   \langle \hat n_1(t \rightarrow + \infty) \rangle &  0 \\
   0 & \langle \hat n_2(t \rightarrow + \infty) \rangle \\
   \end{array} \right) \nn \\
   &=& \left( \begin{array}{c c}
   1 - P_{\rm LZ} &  0 \\
   0 & P_{\rm LZ}  \\
   \end{array} \right).
\eea
So the eigenvalues of the SPDM are directly given by the Landau-Zener tunneling 
rate if the two modes are not coherent. For strong nonlinearities $g$ this is always 
the case and so the left- and the right-hand sides of the figures
\ref{fig-plz-g-a01} and \ref{fig-plz-g-a10} show an excellent agreement except
for a small region around $g = 0$ in figure \ref{fig-plz-g-a10}.
In the non-interacting case ($g = 0$) the dynamics of all atoms is identical and the condensate will be fully coherent at all times. Thus the leading eigenvalue of the SPDM is always equal to unity independent of the Landau-Zener tunneling rate, such that the approximation (\ref{eqn_applzSPDM}) is no longer valid in the non-interacting case.

\section{Semiclassical and adiabatic limit}
\label{sec-limit}

\begin{figure}[t]
\centering
\includegraphics[width=6cm, angle=0]{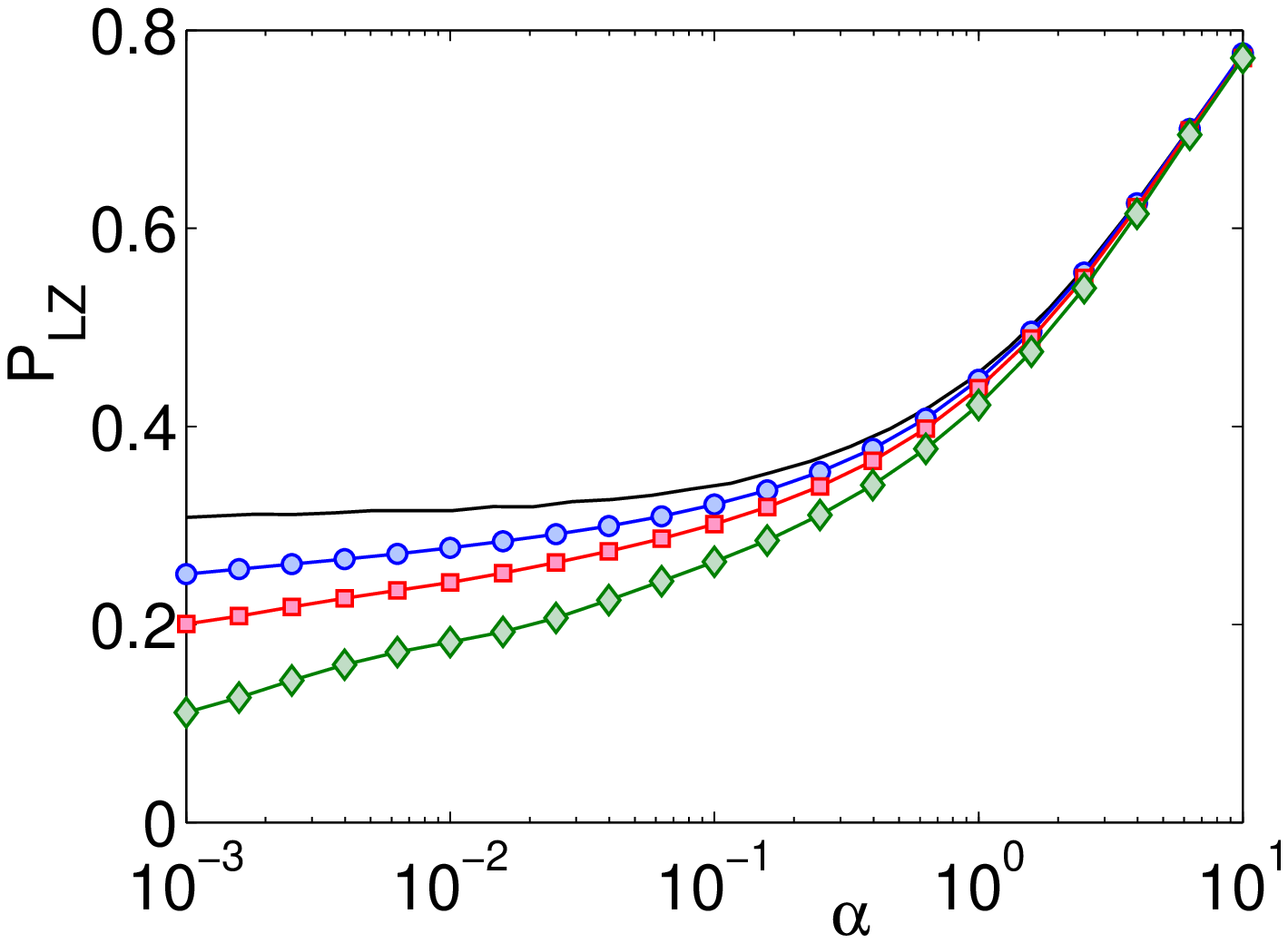}
\includegraphics[width=6cm, angle=0]{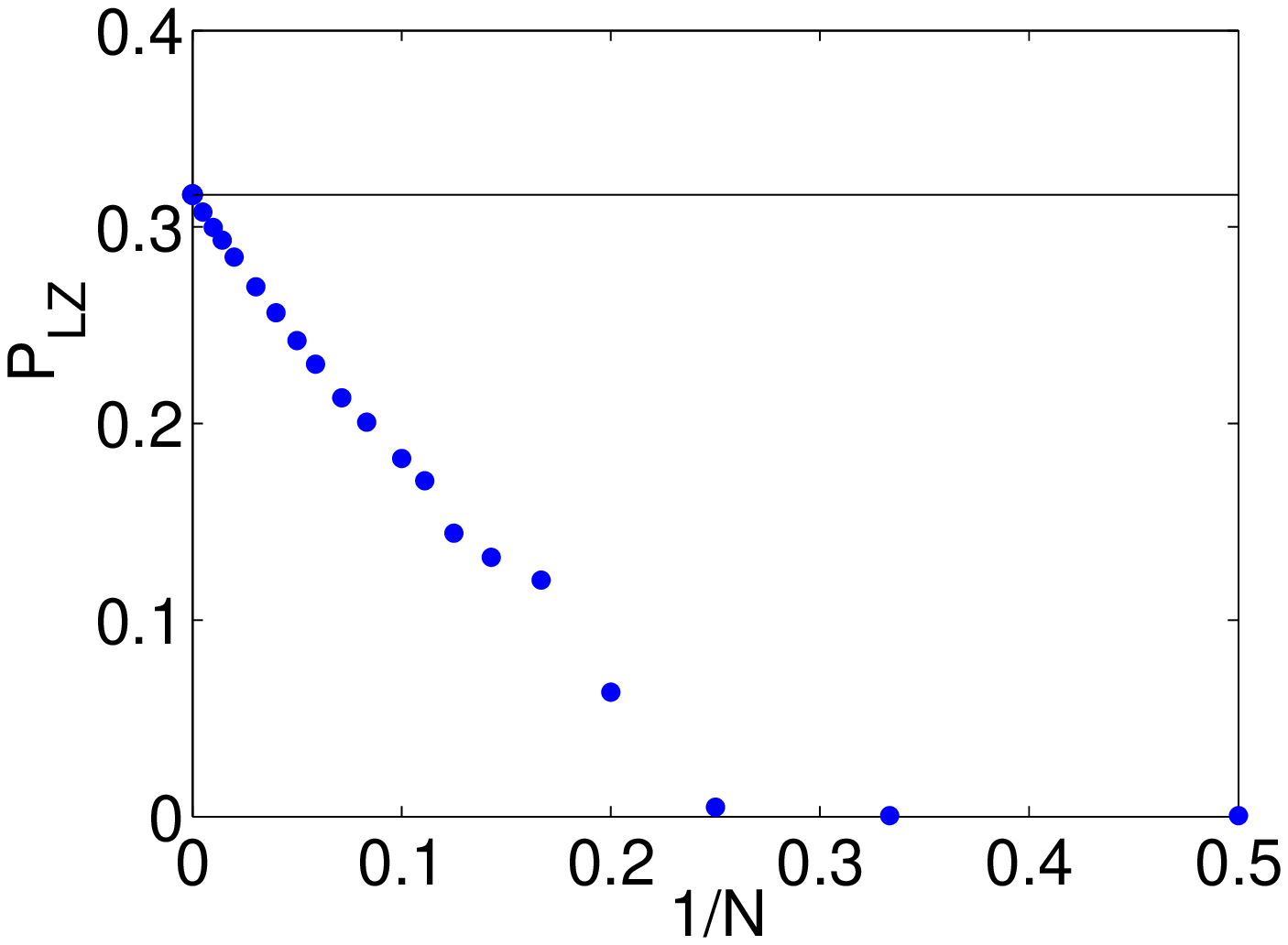}
\caption{\label{fig-plz-mpmf1}
Left panel: Landau-Zener tunneling probability $P_{\rm LZ}$ as a function of $\alpha$
for $g=-5$. Single-trajectory mean-field results (solid black line) are compared to exact 
many-particle results for $N = 10$ (green diamonds), $N=20$ (red squares) and $N=40$ 
(blue circles) particles.  The lines are drawn to guide the eye.
Right panel: Slow convergence to the mean-field limit in the adiabatic regime. The Landau-Zener 
tunneling probability $P_{\rm LZ}$ is plotted as a function of the inverse particle number
for $\alpha=0.01$ and $g=-5$. The black line corresponds to the single-trajectory 
mean-field results which are approached in the limit $1/N \rightarrow 0$.
}
\end{figure}

Having discussed various aspects of the mean-field many-particle correspondence 
in the previous sections, we now investigate the convergence to the mean-field limit 
quantitatively. The left-hand side of figure \ref{fig-plz-mpmf1} compares the 
mean-field Landau-Zener tunneling probability $P_{\rm LZ}(\alpha)$ (\ref{eqn-plz-mf-def})
to the corresponding many-particle results (\ref{eqn-plz-mp-def}) for different particle 
numbers and $g=-5$. While the many-particle dynamics usually converges rapidly 
to the mean-field limit, the occurence of a dynamical instability for $|g| > 2J$ leads to a 
breakdown of adiabaticity for small values of $\alpha$. In this parameter region the 
convergence to the many-particle limit is logarithmically slow. 
This is further illustrated in figure \ref{fig-plz-mpmf1} on the right-hand side, where the 
Landau Zener tunneling probability $P_{\rm LZ}^{\rm mp}$ is plotted 
as a function of the inverse particle number $1/N$. 

Another observation that can be drawn from the numerical data presented in
figure \ref{fig-plz-mpmf1} is that a simple mean-field description gives qualitatively
wrong results in the adiabatic limit of small $\alpha$.
As already discussed in section\ref{sec-mpmf}, the many-particle Landau-Zener tunneling
probability $P_{\rm LZ}^{\rm mp}(\alpha)$ will always tend to zero for 
$\alpha \rightarrow 0$ since the level splittings in the 
many-particle spectrum may become small, but are always non-zero for finite $N$.
Its mean-field counterpart $P_{\rm LZ}^{\rm mf}(\alpha)$ , however, is always 
affected by the appearance of the dynamical instability which destroys 
adiabaticity also for infinitesimally small values of $\alpha$. Consequently, the
Landau-Zener tunneling probability is believed to be non-zero even in this limit.
This difference led to the claim that the adiabatic limit $\alpha \rightarrow 0$
and the semiclassical limit $1/N \rightarrow 0$ do {\it not} commute \cite{Wu06}.
However, this claim is true only for the single-trajectory mean-field description
which assumes a pure condensate at all times, which is obviously no longer
true in the present case.

\begin{figure}[tb]
\centering
\includegraphics[width=6cm, angle=0]{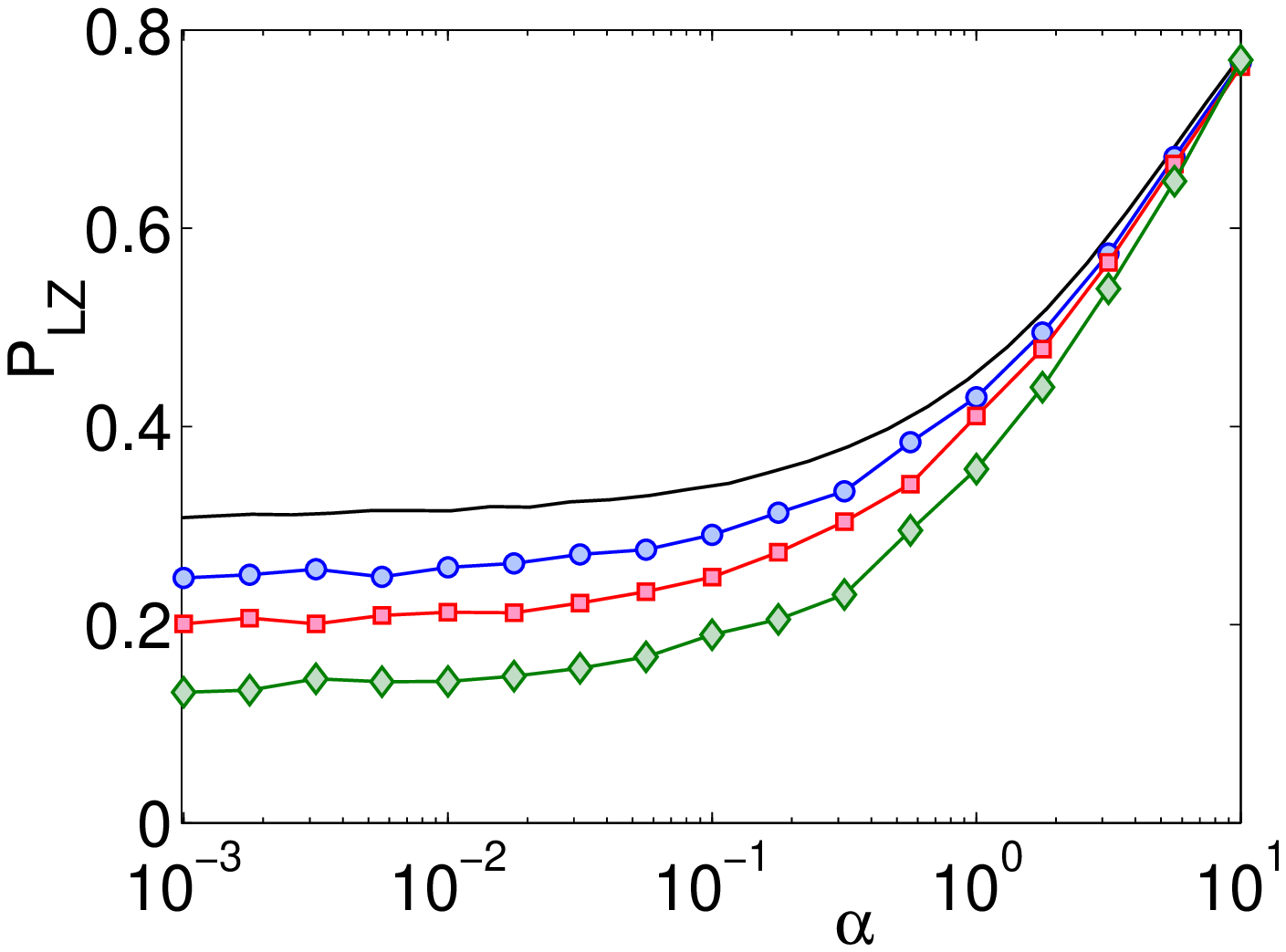}
\includegraphics[width=6cm, angle=0]{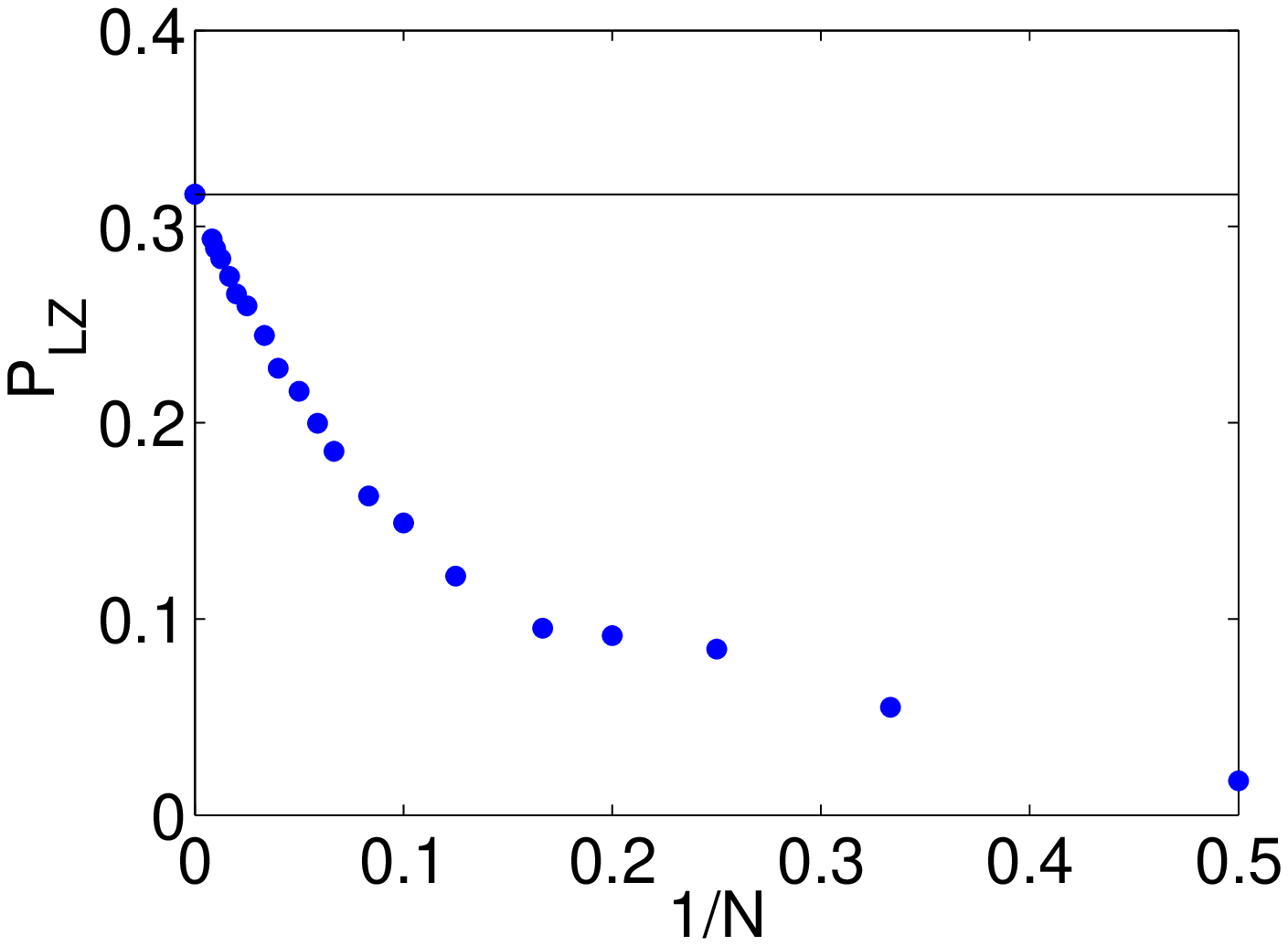}
\caption{\label{fig-plz-ens1}
Left panel: Landau-Zener tunneling probability $P_{\rm LZ}^{\, \rm ens}$ as a function 
of $\alpha$ calculated from a semi-classical ensemble simulation for $g=-5$
 and $N = 10$ (green diamonds), $N=20$ (red squares) and $N=40$ (blue 
 circles) particles. The lines are drawn to guide the eye.
The black line corresponds to the single-trajectory 
mean-field results which are approached in the limit $1/N \rightarrow 0$.
Right panel: The Landau-Zener tunneling probability $P_{\rm LZ}^{\, \rm ens}$ 
as a function of the inverse particle number calculated from a semi-classical 
ensemble simulation for $\alpha=0.01$ and $g=-5$. 
}
\end{figure}

As discussed in the previous section, the proper semiclassial limit of the quantum 
dynamics is a phase space flow rather than a single phase space trajectory. 
This description is valid also if the classical dynamics is unstable and the 
many-particle quantum state deviates from a pure condensate.  
The left-hand side of figure \ref{fig-plz-ens1} shows the Landau-Zener tunneling 
probability $P_{\rm LZ}^{\, \rm ens}(\alpha)$ for different particle numbers calculated 
from the propagation of a semiclassical phase space ensemble as described in section 
\ref{sec-phase}. It is observed that the many-particle results (cf. figure 
\ref{fig-plz-mpmf1}) can be reproduced to a very good approximation 
even for small values of $\alpha$. Thus there is no incommutability
of the adiabatic and semiclassical limits if the latter is interpreted correctly. 
Also the slow convergence to the single-trajectory limit is well described
by the semiclassical phase space approach. The right-hand side of 
figure \ref{fig-plz-ens1} shows  $P_{\rm LZ}^{\, \rm ens}$ as a function
of the inverse particle number $1/N$ for $\alpha = 0.01$ which is well
in  the adiabatic regime. Significant differences to the many-particle results
(cf. figure \ref{fig-plz-mpmf1}) are observed only for very small particle
numbers, $N \leapprox 10$.

\section{Influence of phase noise}
\label{sec-noise}

We finally want to approach the question how an interaction with the environment
affects the transition from quantum-many body to the classical mean-field
dynamics. To this end we consider the Landau-Zener problem subject
to phase noise, which is the dominant influence of the environment provided
that the two condensate modes are held in sufficiently deep trapping
potentials \cite{Angl97,Ruos98}. The many-particle dynamics is then given
by the master equation
\be
  \frac{\rd}{\rd t} \hat \rho  =  -\ri [\hat H,\hat \rho] 
  - \frac{\gamma}{2} \sum_{j = 1,2} \left( \hat n_j^2 \hat \rho + \hat \rho \hat n_j^2 
          - 2 \hat n_j \hat \rho \hat n_j  \right).
\ee 
The effect of phase noise can be included in a single-trajectory mean-field limit 
starting from the dynamics of the Bloch vector (\ref{eqn-gpe-bloch}), whose 
evolution equations are then given by \cite{08mfdecay,08sr4,09srlong}
\bea
  \dot s_x &=& -2 \epsilon s_y - 2 U s_y s_z - \gamma s_x, \nn \\
  \dot s_y &=&  2 J s_z + 2\epsilon  s_x + 2 U s_x s_z  - \gamma s_y, \nn \\
  \dot s_z &=& - 2 J s_y .
  \label{eqn-mf-diss}
\eea
Thus, phase noise leads to transversal relaxation degrading the coherences
$s_x$ and $s_y$ of the two condensate modes. Note that the magnitude of
the Bloch vector is no longer conserved because of this effect.

\begin{figure}[t]
\centering
\includegraphics[width=6cm, angle=0]{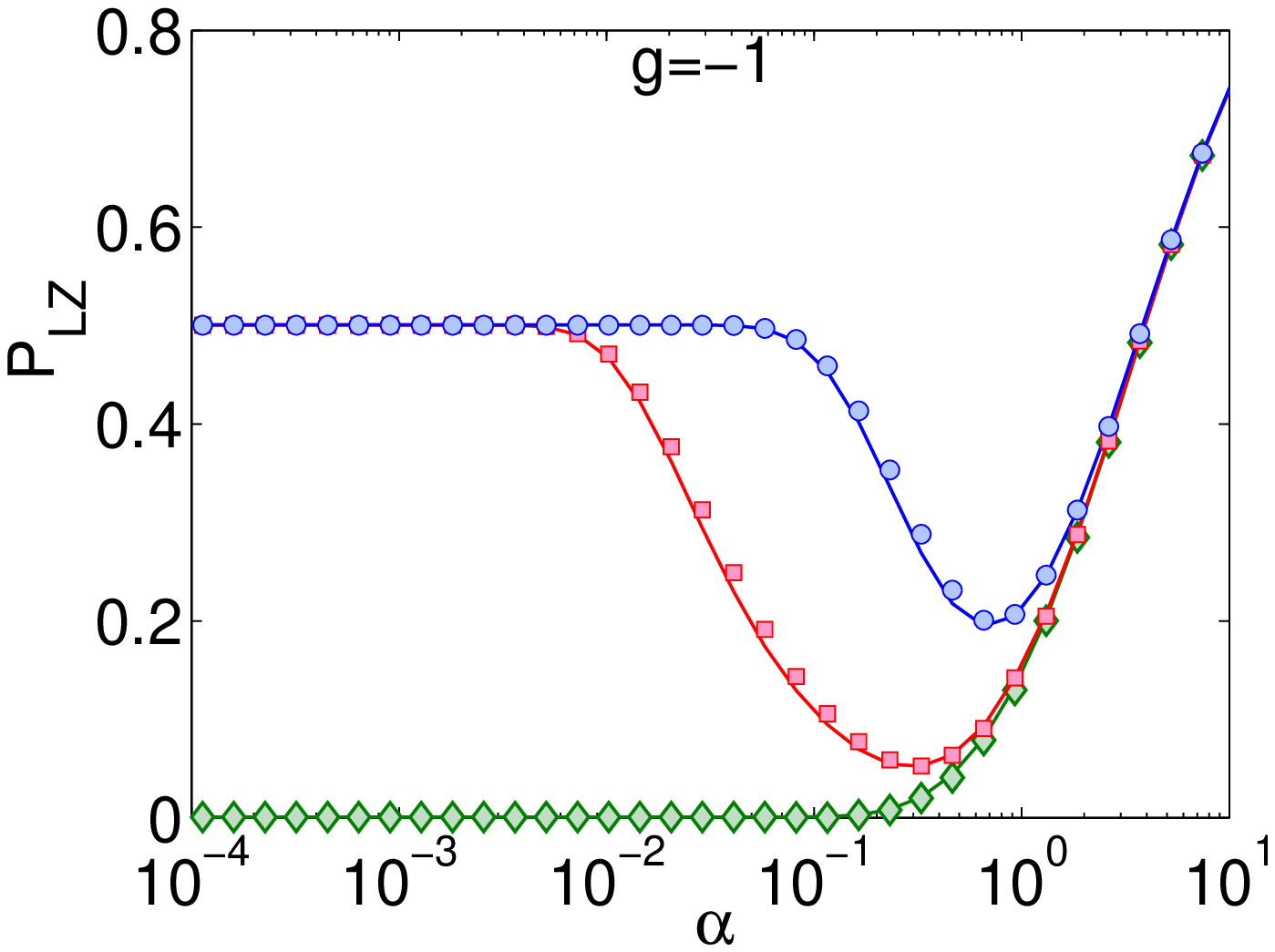}
\includegraphics[width=6cm, angle=0]{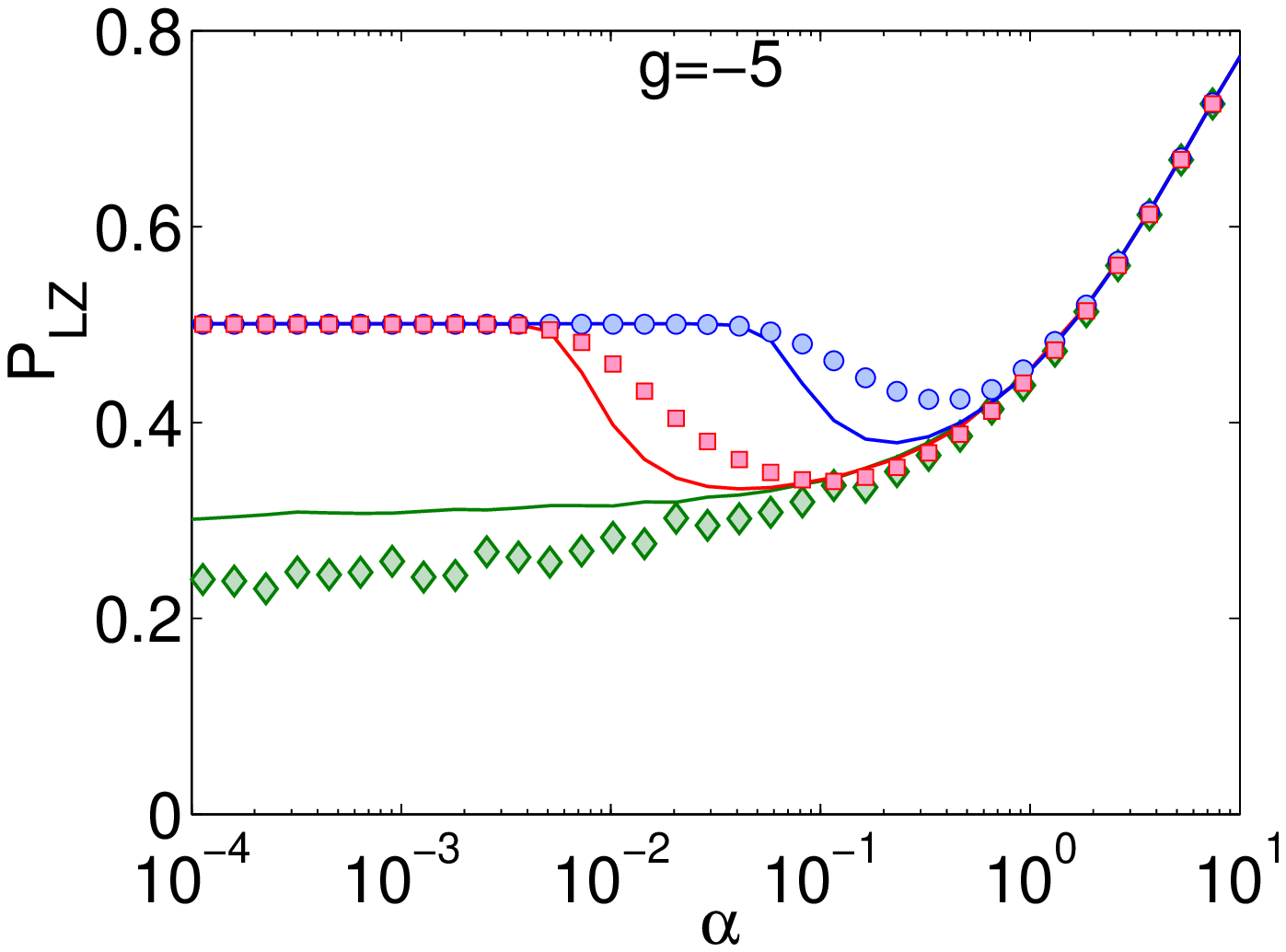}
\caption{\label{fig-plz-diss1}
Landau-Zener tunneling probability $P_{\rm LZ}(\alpha)$ in the
presence of phase noise for $g = -1$ (left) and $g=-5$ (right). 
The strength of the phase noise is chosen as $\gamma = 0$ 
(green diamonds), $\gamma = 0.01$ (red squares) and $\gamma = 0.1$
(blue circles). Mean-field results (solid lines) are compared to 
many-particle results  $N=40$ particles (symbols). 
}
\end{figure}

The resulting Landau-Zener tunneling probabilities are plotted in 
figure \ref{fig-plz-diss1} as a function of $\alpha$ for different values of the
noise strength $\gamma$. It is observed that phase noise has an important
effect only for small values of $\alpha$, where it drives the system to a 
completely mixed state with equal population in both wells so that
$P_{\rm LZ} = 1/2$. On the contrary, almost no consequences are observed 
for fast parameter sweeps. In this case, the tunneling time during which the  
atoms are delocalized is so short that phase noise cannot affect the 
dynamics. 
The transition to the incoherent regime occurs when the time scale of
the noise $\gamma^{-1}$ is smaller than the tunneling time which is
roughly given by $\alpha^{-1}$. Therefore the sweep is incoherent
such that  $P_{\rm LZ} = 1/2$ if
\be
   \alpha \leapprox \gamma,
\ee
while the interaction strength $g$ has a minor effect only.

Comparing mean-field and many-particle results, significant differences are
observed for very small values of $\alpha$ and $g = -5$ in the non-dissipative case 
$\gamma = 0$, which has been discussed in detail above. In addition we 
note that already a small amount of phase noise is sufficient to remove
these differences. For $\alpha \rightarrow 0$ and $\gamma \neq 0$ the
mean-field approximation (\ref{eqn-mf-diss}) correctly predicts the 
transition to a completely mixed state with $P_{\rm LZ} = 1/2$.
Furthermore, significant differences are observed for $g = -5$ and intermediate 
values of $\alpha$. In this case the many-particle quantum state is no longer 
a pure BEC but rather strongly number squeezed as discussed above.
This state is more easily driven to a completely mixed state by phase noise
than a pure BEC, a process which certainly cannnot be described by the 
simple single-trajectory mean-field approximation.

Finally, these results suggest that Landau-Zener sweeps may actually 
be used as a probe of decoherence in systems of ultracold atoms 
(cf. also \cite{Wubs06}). A measurement of the transition point to the
incoherent regime where $P_{\rm LZ} = 1/2$ gives an accurate
quantitative estimate of the noise strength $\gamma$ with a fairly 
simple experiment.

\section{Conclusion and Outlook}

In the present paper we have presented an analysis of nonlinear Landau-Zener 
tunneling between two modes in quantum phase space. It was shown that 
adiabaticity breaks down if the interaction strength $g=UN$ exceeds the critical
value $2J$-- the Landau-Zener tunneling probability does not vanish even for an 
extremely slow variation of the system parameter.
This phenomenon can be understood by the disappearance of adiabatic eigenstates 
in an inverse bifurcation in the mean-field approximation. Within the full 
many-particle description, the breakdown of adiabaticity results from the 
occurrence of diabatic avoided crossings, where the level separation vanishes 
exponentially with the number of particles.

The correspondence of the quantum dynamics and the `classical' mean-field 
approximation has been discussed in detail. The many-particle and the 
mean-field Landau-Zener tunneling probability show an excellent agreement, 
because quantum fluctuations of the populations are small.
In contrast, there is no fixed phase relation between the two modes,
which certainly goes beyond the simple Bogoliubov mean-field theory.
An improved classical approximation using phase space ensembles can 
describe the depletion of the condensate mode and the loss of phase 
coherence as well as number squeezing $\xi_N^2$ of the final state. 
Yet temporal revivals of this coherence are genuine many-particle
effects and cannot be described  classically. Thus, the spectroscopically
relevant squeezing parameter $\xi_S^2$ is not reproduced by the ensemble simulation. However, 
the timescale for the occurence of these revivals and accordingly of the spectroscopical squeezing 
depends linearly on the particle number. For realistic setups, this is way to long compared to 
decoherence and phase noise rates. Before the sytem reaches the squeezed state, nearly all 
coherences are already lost.

In the last section, we have studied how the dynamics depends on the number of 
particles $N$ and 
compare our results to the discrete Gross-Pitaevskii equation that describes 
the dynamics in the limit $N \rightarrow \infty$. We show that the contradiction 
between the mean-field prediction and the exact many-particle transition rate 
in the adiabatic regime is no longer present in the phase space approach, 
and must therefore be considered as an artifact of the single-trajectory 
description. 
These results demonstrate the power of the phase space approach.
However, in order to reproduce true quantum features such
as quantum beats semiclassically, a more refined treatment is 
necessary. Semiclassical coherent state propagators have been 
studied intensively in single particle quantum mechanics in the
limit $\hbar \rightarrow 0$ \cite{Bara01}. An extension to the 
mean-field limit of quantum many-body system must be based
on the $SU(M)$ phase space discussed in the present paper. 
In this case the particle number $N$ is a number and not an
operator and $1/N$ will serve as a proper semiclassical 
parameter. However, a numerical calculation for realistic particle numbers 
based on these methods, taking into account all relevant phase information 
between different trajectories, is as hard as the original quantum problem.

Furthermore, we show that already the presence of a small amount of phase 
noise is sufficient to introduce enough decoherence to make the system 'classical',
so that the many-particle dynamics is well reproduced with a simple single-trajectory 
mean-field approach. Finally we have argued that a measurement of the transition 
to an incoherent Landau-Zener sweep could be used as a sensitive probe of 
decoherence.

\section*{Acknowledgements}

This work has been supported by the German Research Foundation (DFG) through the 
research fellowship program (grant number WI 3415/1-1) and the Graduiertenkolleg 792
as well as the Studienstiftung des deutschen Volkes. We thank M.~Wubs for valuable
comments.

\end{document}